# Tree Automata for Extracting Consensus from Partial Replicas of a Structured Document

**Maurice Tchoupé Tchendji, Milliam M. Zekeng Ndadji**

Department of Mathematics and Computer Science, Faculty of Sciences, University of Dschang, Dschang, Cameroon
Email: ttchoupe@yahoo.fr, maurice.tchoupe@univ-dschang.org, ndadjimaxime@yahoo.fr

## Abstract

In an asynchronous cooperative editing workflow of a structured document, each of the co-authors receives in the different phases of the editing process, a copy of the document to insert its contribution. For confidentiality reasons, this copy may be only a partial replica containing only parts of the (global) document which are of demonstrated interest for the considered co-author. Note that some parts may be a demonstrated interest over a co-author; they will therefore be accessible concurrently. When it's synchronization time (e.g. at the end of an asynchronous editing phase of the process), we want to merge all contributions of all authors in a single document. Due to the asynchronism of edition and to the potential existence of the document parts offering concurrent access, conflicts may arise and make partial replicas unmergeable in their entirety: they are inconsistent, meaning that they contain conflictual parts. The purpose of this paper is to propose a merging approach said by consensus of such partial replicas using tree automata. Specifically, from the partial replicas updates, we build a tree automaton that accepts exactly the consensus documents. These documents are the maximum prefixes containing no conflict of partial replicas merged.

## Keywords

Structured Documents, Workflow of Cooperative Edition, Merging Partial Replicas, Conflict, Consensus, Tree Automata, Automata Product, Lazy Evaluation

## 1. Introduction

A significant proportion of documents handled and/or exchanged by applications has a regular structure defined by a grammatical model (DTD: *Document Type Definition*, schema [1]): they are called *structured documents*. The ever-increasing power of communication networks in terms of throughput and security, as well as efficiency is concerned, has revolutionized the way of editing such documents. Indeed, to the classical model of an author, editing locally and autonomously his document, was added the (asynchronous) cooperative editing in which, several authors located on geographically distant sites, coordinate to edit asynchronously a same structured document (**Figure 1**): it is an asynchronous cooperative editing workflow.

In such editing workflows (**Figure 2**), the desynchronized editing phases in which each co-author edits on his site his copy of the document, alternate with the synchronization-redistribution phases in which the different contributions (local replicas) are merged (on a dedicated site) into a single document, which is then redistributed to the various co-authors for the continuation of the edition. This pattern is repeated until the document is completely edited.

In the literature, there are several cooperative editing systems offering, for some, a concurrent collaborative edition of the same document (Etherpad [2], Google Docs [3], Fidus Writer [4], …), or on the other hand, a truly distributed and asynchronous edition (Wikis [5] [6], Git [7] [8], …) in which the co-authors work on replicas of the document; replication techniques as well as reconciliation strategies must then be addressed. If the collectively edited document is structured, it may in some cases be desirable for reasons of confidentiality, for example, a co-author has access only to certain information, meaning that he only has access to certain parts of the document belonging to certain given types (*sorts*[1]) of the document model. Thus, the replica $t_i$ edited by co-author $c_i$ from the site $i$ may be only a *partial replica* of the (global[2]) document $t$, obtained via a *projection operation*, which conveniently eliminates from global document *t* parts which are not accessible to the co-author in question. We call "*view*" of a co-author, the set of *sorts* that he can access [9]. **Figure 1** is an illustration of such a cooperative edition in which, the edition and the merging of the (global) document in conformity to the (global) model *G* of documents are perform on site 1; while on site 2 and 3, other co-authors perform the edition of the partial replicas in accordance with projected models of documents $G_1$ and $G_2$ obtained from the global model *G*.

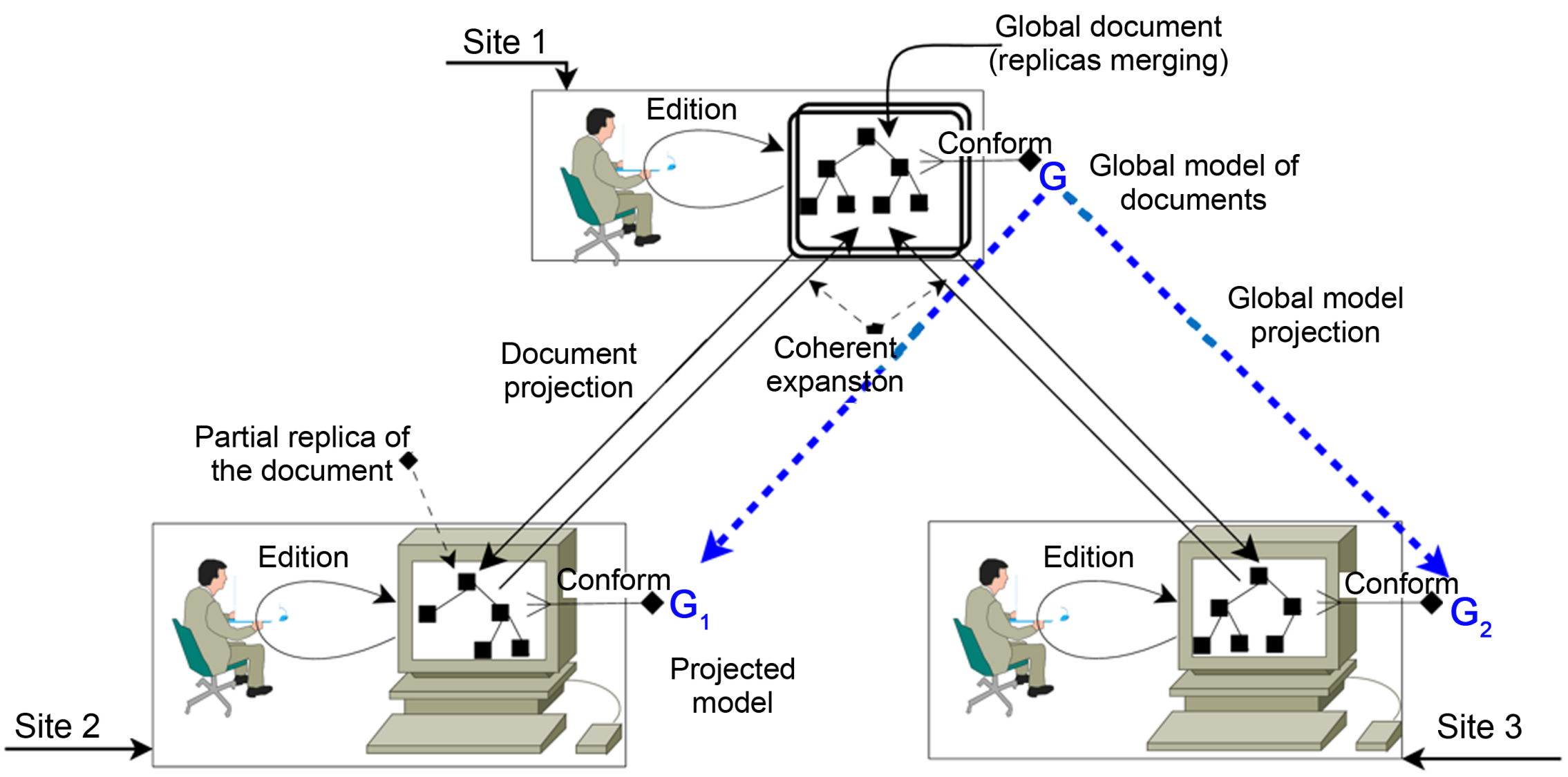

**Figure 1.** The desynchronized cooperative editing of partial replicas of a structured document.

[1]A *sort* is a datum used to define the structuring rules (syntax) in a document model. Example: a *non-terminal symbol* in a context free grammar, an *ELEMENT* in a DTD.

[2]We designate by *global document* or simply *document* when there is no ambiguity, the document including all parts.

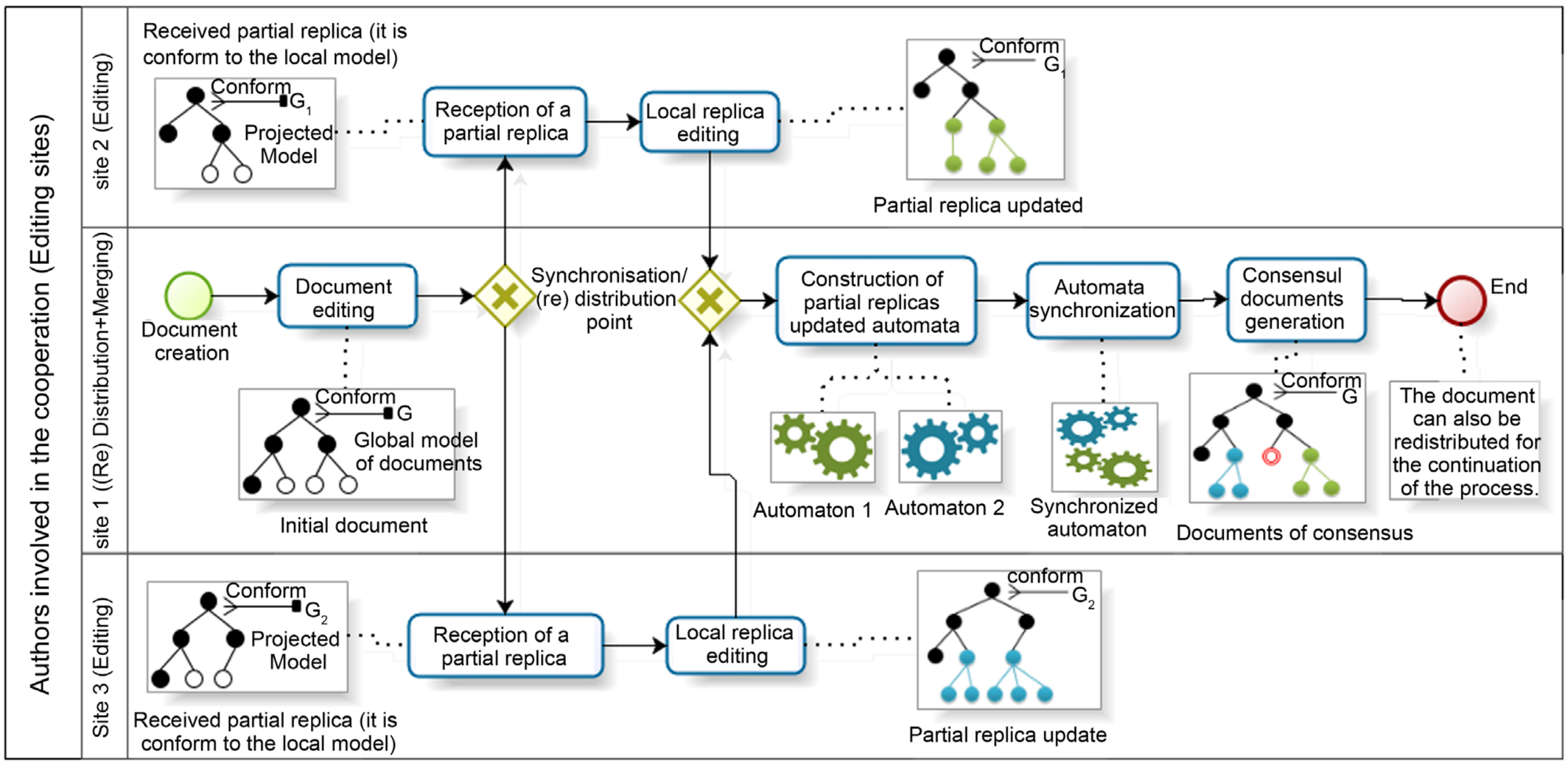

**Figure 2.** An orchestration diagram of an asynchronous cooperative editing workflow.

When asynchronous local editions are done on partial replicas, it can be assumed that each co-author has on his site a local document model guiding him in his edition. This local model can help to ensure that for any update $t_i^{maj}$ of a partial replica $t_i$ (conform to the considered (local) model), there is at least one document *t* conform to the global model so that $t_i^{maj}$ is a partial replica of *t*: for this purpose, the local model should be coherent towards the global one[3]. Thus, because of the asynchronism of the editing, the only inconsistencies that we can have when the synchronization time arrives are those from the concurrent edition of the same node[4] (in the point of view of the global document) by several co-authors: the partial replicas concerned are said to be in conflict. This paper proposes an approach of detection and resolution of such conflicts by *consensus* during the synchronisation-redistribution phase, using a tree automaton said of consensus, to represent all documents that are the consensus of competing editions realised on the different partial replicas.

[3]Intuitively, a local model of document is *coherent* towards a global model if any (partial) document $t_i$ that is conform to it, is the partial replica of at least one (global) document *t* conform to the global model.

[4]Manipulated documents are structured; they can be intentionally represented by a tree. Intuitively, a node is an identifiable part in the document (a section, a subsection, an image, a table, …): it is the instance of a sort.

A structured document $t$ is intentionally represented by a tree that possibly contains buds[5] [9] (see Figure 3). Intuitively, synchronizing or merging consensually the updates $t_1,\cdots,t_n$ of $n$ partial replicas of a document $t$, consists in finding a document $t_c$ conform to the global model, integrating all the nodes of $t_i$ not in conflict and in which, all the conflicting nodes are replaced by buds. Consensus documents are therefore the largest prefixes without conflicts in merged documents. The algorithm of consensual merging presented in this paper is an adaptation of the fusion algorithm presented in [9] which does not handle conflicts. Technically, the process for obtaining the documents forming part of the consensus is: 1) for each update $t_i^{maj}$ of a partial replica $t_i$, we associate a tree automaton with *exit states* $\mathcal{A}^{(i)}$ recognizing the trees (conform to the global model) for which $t_i^{maj}$ is a projection [9]. 2) The consensual automaton $\mathcal{A}_{(sc)}$ generating the consensus documents is obtained by performing a *synchronous product* of the automata $\mathcal{A}^{(i)}$ with a commutative and associative operator noted $\otimes$ that we define. It is such that: $\mathcal{A}_{(sc)} = \otimes \mathcal{A}^{(i)}$. 3) It only remains to generate the set of trees (or those most representative) accepted by the automaton $\mathcal{A}_{(sc)}$, to obtain the consensus documents.

In the subsequent sections, after the presentation (Section 2) of some concepts and definitions related to the cooperative editing and tree automata, we expose (Section 3) the construction process of the operator $\otimes$ and a proof of correction of the algorithm proposed for its implementation. The Section 4 is devoted to the conclusion. In the appendices, we fully unfold the example introduced in Section 3 highlighting the major concepts outlined in this paper (Appendix A), as well as some screenshots of an asynchronous cooperative editor prototype operating in a distributed environment that we have developed for the experimental purposes of the algorithms described in this paper (Appendix B).

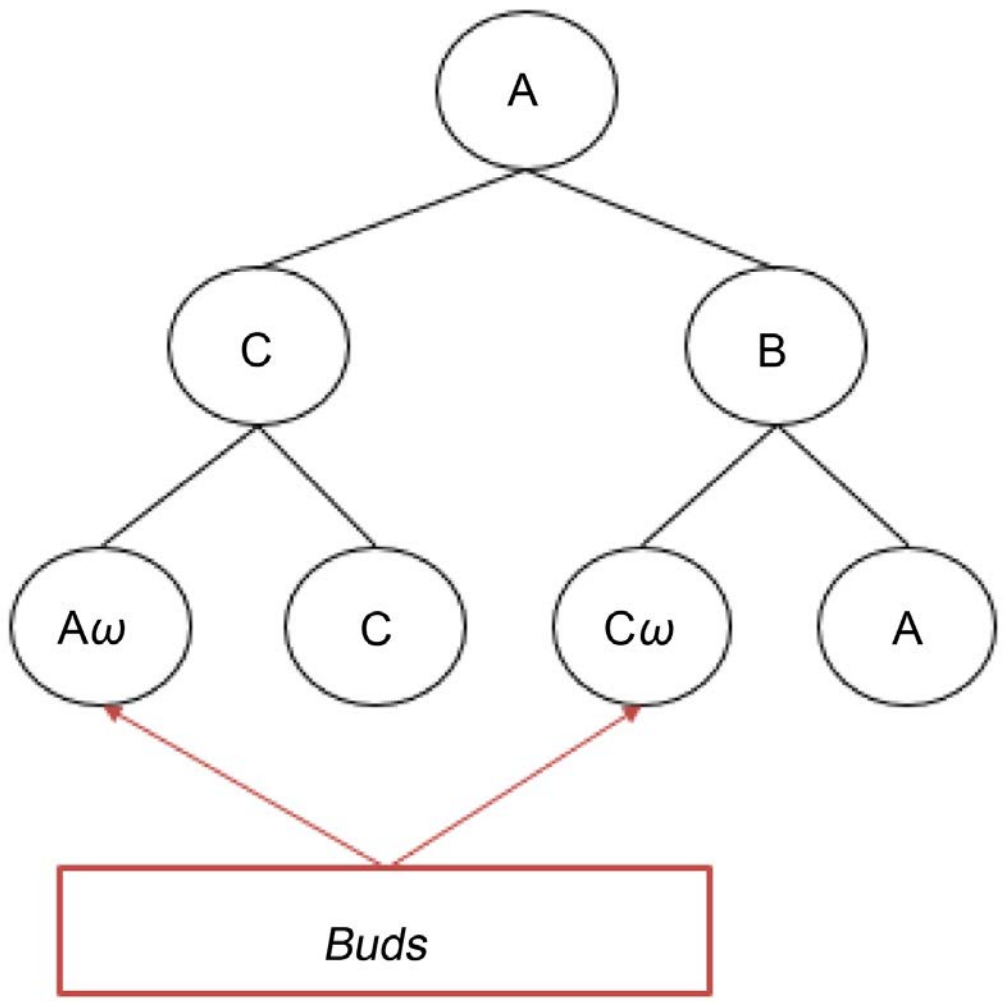

Figure 3. An intentional representation of a document containing buds.

[5]A *bud* is a leaf node of a tree indicating that an edition must be done at that level in the tree. Edit a bud consists to replace it by a sub-tree using the productions of the grammar of the document.

## 2. Structured Cooperative Edition and Notion of Partial Replication

### 2.1. Structured Document, Edition and Conformity

In the XML[6] community, the document model is typically specified using a Document Type Definition (DTD) or a XML Schema [1]. It is shown that these DTD are equivalent to (regular) grammars with special characteristics called *XML grammars* [10]. The (context free) grammars are therefore a generalization of the DTD and on the basis of the studies they have undergone, mainly as formal models for the specification of programming languages, they provide an ideal framework for formal study of the transformations involved in XML technologies. That's why we use them in our work as tools for specifying the structure of documents.

We will therefore represent the abstract structure of a structured document by a tree and its model by an abstract context free grammar; a valid structured document will then be a derivation tree for this grammar. A context free grammar defines the structure of its instances (the documents that are conform to it) by means of the productions. A production, generally denoted $p: X_0 \to X_1 \cdots X_n$ is comparable in this context, to a structuring rule which show how the symbol $X_0$ located in the left part of the production is divided into a sequence of other symbols $X_1 \cdots X_n$ located on its right side. More formally.

**Definition 1.** *A* ***abstract context free grammar*** *is given by* $\mathbb{G} = (\mathcal{S}, \mathcal{P}, \mathcal{A})$ *composed of a finite set* $\mathcal{S}$ *of* ***grammatical symbols*** *or* ***sorts*** *corresponding to the different* ***syntactic categories*** *involved, a particular grammatical symbol* $\mathcal{A} \in \mathcal{S}$ *called* ***axiom***, *and a finite set* $\mathcal{P} \subseteq \mathcal{S} \times \mathcal{S}^*$ *of* ***productions***. *A production* $P = \left(X_{P(0)}, X_{P(1)}, \cdots, X_{P(n)}\right)$ *is denoted* $P: X_{P(0)} \to X_{P(1)} \cdots X_{P(n)}$ *and* $|P|$ *denotes the length of the right hand side of P. A production with the symbol X as left part is called a X-production.*

For certain treatments on trees (documents) it is necessary to designate precisely a particular node. Several indexing techniques exist, among them, the so-called Dynamic Level Numbering [11] based on identifiers with variable lengths inspired by the *Dewey* decimal classification (see **Figure 4**). According to this indexing system, a tree can be defined as follows:

**Definition 2.** *A* ***tree*** *whose nodes are labelled in an alphabet* $\mathcal{S}$ *is a partial map* $t: \mathbb{N}^* \to \mathbf{S}$ *whose domain* $Dom(t) \subseteq \mathbb{N}^*$ *is a prefix closed set such that, for all* $u \in Dom(t)$ *the set* $\{i \in \mathbb{N} \mid u \cdot i \in Dom(t)\}$ *is a not empty interval of integers* $[1, \cdots, n] \cap \mathbb{N}$ ($\varepsilon \in Dom(t)$ *is the root label*); *the integer n is the* ***arity*** *of the node whose address is u.* $t(w)$ *is the value* (*label*) *of the node in t whose address is w. If* $t_1, \cdots, t_n$ *are trees and* $a \in \mathbf{S}$, *we denote* $t = a[t_1, \cdots, t_n]$ *the tree t of domain* $Dom(t) = \{\varepsilon\} \cup \{i \cdot u \mid 1 \le i \le n, u \in Dom(t_i)\}$ *with* $t(\varepsilon) = a$ *and* $t(i \cdot u) = t_i(u)$.

Let *t* be a document and $\mathbb{G} = (\mathcal{S}, \mathcal{P}, \mathcal{A})$ a grammar. *t* is a derivation tree for $\mathbb{G}$ if its root node is labelled by the axiom $A$ of $\mathbb{G}$, and if for all internal node $n_0$ labelled by the sort $X_0$, and whose sons $n_1,\cdots,n_n$, are respectively labelled by the sorts $X_1,\cdots,X_n$, there is one production $P \in \mathcal{P}$ such that, $P: X_0 \to X_1 \cdots X_n$ and $|P| = n$. It is also said in this case that $t$ belongs to the language generated by $\mathbb{G}$ from the symbol $A$ and it is denoted $t \in \mathscr{L}(\mathbb{G}, A)$ or $t \therefore \mathbb{G}$.

[6]XML is the acronym for *Extensible Markup Language*.

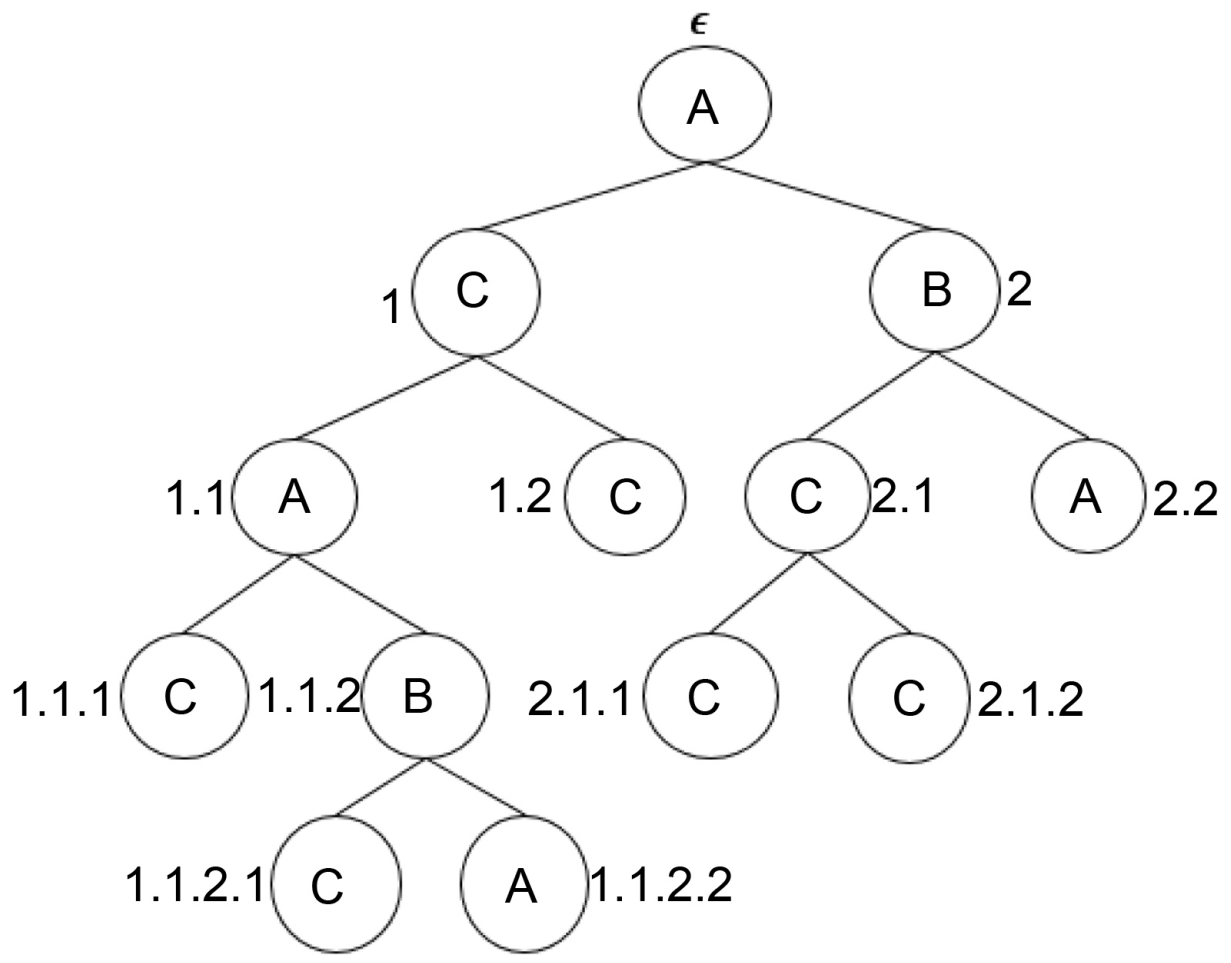

**Figure 4.** Example of an indexed tree.

There is a bijective correspondence between the set of derivation trees of one grammar and all its Abstract Syntax Tree (*AST*). In an AST, nodes are labelled by the names of the productions.

**Definition 3.** *The set* $AST(\mathbb{G}, X)$ *of* ***abstract syntax trees*** *according to the grammar* $\mathbb{G}$ *associated with grammatical symbol X consists of trees in the form* $P[t_1,\cdots,t_n]$ *where P is a production such that* $X = X_{P(0)}$, $n = |P|$ *and* $t_i \in AST(\mathbb{G}, X_i), X_i = X_{P(i)}$ *for all* $1 \le i \le n$.

*AST* are used to show that a given tree labelled with grammatical symbols is an instance of a given grammar.

A structured document being edited is represented by a tree containing *buds* (or *open nodes*) which indicate in a tree, the only places where editions (*i.e.* updates) are possible[7]. Buds are typed; a *bud of sort X* is a leaf node labelled by $X_\omega$: it can only be edited (*i.e* extended to a sub-tree) by using a *X-production* of the form $P: X \to X_1 \cdots X_n$ which have as effect, 1) the replacement of $X_\omega$ labelled bud by a $P$ labelled node and, 2) the creation of $n$ buds labelled respectively by $X_{i\omega}, 1 \le i \le n$. Thus, a structured document being edited and that have the grammar $\mathbb{G} = (S, \mathcal{P}, A)$ as model, is a derivation tree for the extended grammar $\mathbb{G}_\Omega = (S \cup S_\omega, \mathcal{P} \cup S_\Omega, A)$ obtained from $\mathbb{G}$ as follows: for all sort $X$, we not only add in the set $S$ of sorts a new sort $X_\omega$, but we also add a new $\varepsilon$-production $X_\Omega : X_\omega \to \varepsilon$ in the set $\mathcal{P}$ of productions; so we have: $S_\omega = \{X_\omega, X \in S\}$ and $S_\Omega = \{X_\Omega : X_\omega \to \varepsilon, X_\omega \in S_\omega\}$.

[7]We are interested in this paper only in the *positive edition*, based on a partial optimistic replication [12] of edited documents; In fact, the published documents are only increasing: there is no erasure possible as soon as a synchronization has been performed.

When we look at the productions of a grammar, we can notice that each sort is associated with a set of productions. From this point of view, therefore, we can consider a grammar as an application

$$gram : symb \rightarrow \left[\left(prod, [symb]\right)\right] \quad (1)$$

which associates to each sort a list of pairs formed by a production name and the list of sorts in the right hand side of this production. Such an observation suggests that a grammar can be interpreted as a (descending) tree automaton that can be used for recognition or for the generation of its instances.

**Definition 4.** *A (descending)* ***tree automaton*** *defined on* $\Sigma$ *is a quadruplet* $\mathcal{A} = (\Sigma, Q, R, q_0)$ *of a set* $\Sigma$ *of symbols; its elements are the labels of the nodes of the trees to be generated (or recognized), a set Q of states, a particular state* $q_0 \in Q$ *called initial state, and a finite set* $R \subseteq Q \times \Sigma \times Q^*$ *of transitions.*

- An element of $R$ is denoted $q \rightarrow (\sigma, [q_1, \cdots, q_n])$ or in an equivalent way $q \xrightarrow{\sigma} (q_1, \cdots, q_n)$: intuitively, $[q_1, \cdots, q_n]$ is the list of states accessible from the $q$ state by crossing a transition labelled $\sigma$.
- If $q \xrightarrow{\sigma_1} (q_1^1, \cdots, q_{n_1}^1), \cdots, q \xrightarrow{\sigma_k} (q_1^k, \cdots, q_{n_k}^k)$ denotes the set of transitions associated to the state $q$, we denote $next\ q = \left[\left(\sigma_1, \left[q_1^1, \cdots, q_{n_1}^1\right]\right), \cdots, \left(\sigma_k, \left[q_1^k, \cdots, q_{n_k}^k\right]\right)\right]$ the list that consists of pairs $\left(\sigma_i, \left[q_1^i, \cdots, q_{n_i}^i\right]\right)$. A transition of the form $q \rightarrow (\sigma, [\,])$ is called ***final transition*** and a state possessing this transition is a ***final state***.

One can interpret a grammar $\mathbb{G} = (S, \mathcal{P}, A)$ as a (descending) tree automaton [13] $\mathcal{A} = (\Sigma, Q, R, q_0)$ considering that: 1) $\Sigma = \mathcal{P}$ is the type of labels of the nodes forming the tree to recognize. 2) $Q = S$ is the type of states and, 3) $q \rightarrow (\sigma, [q_1, \cdots, q_n])$ is a transition of the automaton when the pair $(\sigma, [q_1, \cdots, q_n])$ appears in the list $(gram\ q)$[8]. We note $\mathcal{A}_{\mathbb{G}}$ the tree automaton derived from $\mathbb{G}$.

To obtain the set $AST_{\mathcal{A}}$ of *AST* generated by a tree automaton $\mathcal{A}$ from an initial state $q_0$, you must: 1) Create a root node *r*, associate the initial state $q_0$ and add it to the set $AST_{\mathcal{A}}$ initially empty. 2) Remove from $AST_{\mathcal{A}}$ an AST *t* under construction *i.e* with at least one leaf node *node* unlabelled. Let *q* be the state associated to *node*. For all transition $q \xrightarrow{\sigma} (q_1, \cdots, q_n)$ of $\mathcal{A}$, add in $AST_{\mathcal{A}}$ the trees $t'$ which are replicas of *t* in which the node *node* has been substituted by a node $node'$ labelled $\sigma$ and possessing *n* (unlabelled) sons, each associated to a (distinct) state $q_i$ of $[q_1, \cdots, q_n]$. 3) Iterate step (2) until you obtain trees with all the leaf nodes labelled (they are consequently associated to the final states of $\mathcal{A}$): these are *AST*. We note $\mathcal{A} \vDash t \rhd q$ the fact that the tree automaton $\mathcal{A}$ accepts the tree *t* from the initial state *q*, and $\mathscr{L}(\mathcal{A}, q)$ (tree language) the set of trees generated by the automaton $\mathcal{A}$ from the initial state *q*. Thus, $(\mathcal{A} \vDash t \rhd q) \Leftrightarrow (t \in \mathscr{L}(\mathcal{A}, q))$.

As for automata on words, one can define a synchronous product on tree automata to obtain the automaton recognizing the intersection, the union, …, of regular tree languages [13]. We introduce below the definition of the synchronous product of $k$ tree automata whose adaptation will be used in the next section for the derivation of the consensual automaton.

[8]Reminder: *gram* is the application obtained by abstraction of $\mathbb{G}$ and have as type: $gram : symb \rightarrow \left[\left(prod, [symb]\right)\right]$.

**Definition 5.** ***Synchronous product of k automata***:

Let $\mathcal{A}_1 = \left(\Sigma, Q^{(1)}, R^{(1)}, q_0^{(1)}\right), \cdots, \mathcal{A}_k = \left(\Sigma, Q^{(k)}, R^{(k)}, q_0^{(k)}\right)$ be $k$ tree automata. The synchronous product of these $k$ automata $\mathcal{A}_1 \otimes \cdots \otimes \mathcal{A}_k$ denoted $\otimes_{i=1}^{k} \mathcal{A}^{(i)}$, is the automaton $\mathcal{A}_{(sc)} = (\Sigma, Q, R, q_0)$ defined as follows: 1) Its states are vectors of states : $Q = Q^{(1)} \times \cdots \times Q^{(k)}$; 2) Its initial state is the vector formed by the initial states of the different automata : $q_0 = \left(q_0^{(1)}, \cdots, q_0^{(k)}\right)$; 3) Its transitions are given by:

$$\left(q^{(1)}, \cdots, q^{(k)}\right) \xrightarrow{a} \left(\left(q_1^{(1)}, \cdots, q_1^{(k)}\right), \cdots, \left(q_n^{(1)}, \cdots, q_n^{(k)}\right)\right)$$
$$\Leftrightarrow \left(q^{(i)} \xrightarrow{a} \left(q_1^{(i)}, \cdots, q_n^{(i)}\right) \quad \forall i, 1 \le i \le k\right)$$

## 2.2. Notions of View, Projection, Reverse Projection and Merging

### 2.2.1. View, Associated Projection and Merging

The derivation tree giving the (global) representation of a structured document edited in a cooperative way makes visible the set of grammatical symbols of the grammar that participated in its construction. As we mentioned in Section 1 above, for reasons of confidentiality (accreditation degree), a co-author manipulating such a document will not necessarily have access to all of these grammatical symbols; only a subset of them can be considered relevant for him: it is his *view*. A view $\mathcal{V}$ is then a subset of grammatical symbols ($\mathcal{V} \subseteq S$).

A partial replica of $t$ according to the view $\mathcal{V}$, is a partial copy of $t$ obtained by deleting in $t$ all the nodes labelled by symbols that are not in $\mathcal{V}$. Figure 5 shows a document $t$ (center) and two partial replicas $t_{\mathcal{V}_1}$ (left) and $t_{\mathcal{V}_2}$ (right) obtained respectively by projections from the views $\mathcal{V}_1 = \{A, B\}$ and $\mathcal{V}_2 = \{A, C\}$.

Practically, a partial replica is obtained via a *projection* operation denoted $\pi$. We therefore denote $\pi_{\mathcal{V}}(t) = t_{\mathcal{V}}$ the fact that $t_{\mathcal{V}}$ is a partial replica obtained by projection of $t$ according to the view $\mathcal{V}$.

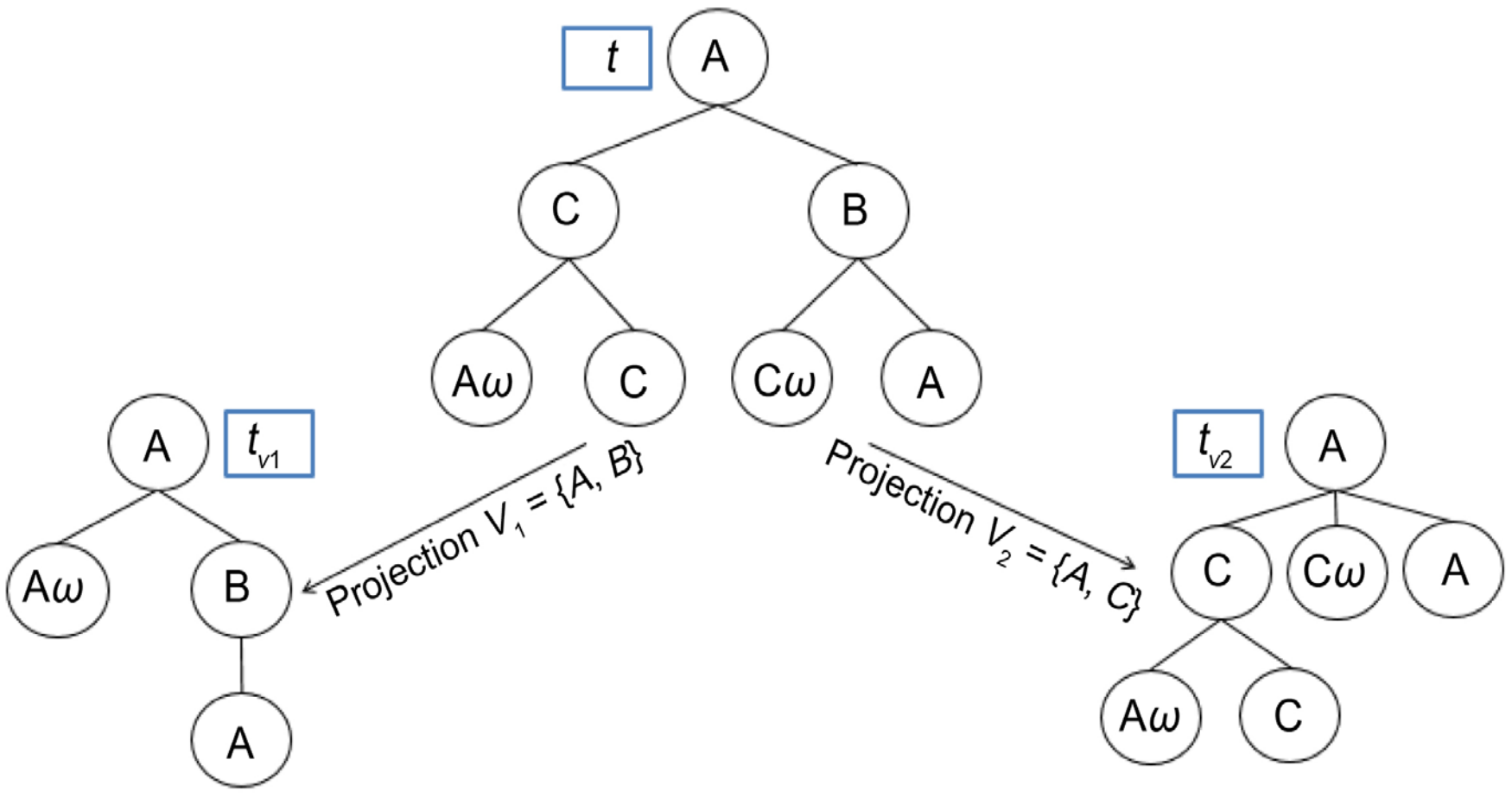

Figure 5. Example of projections made on a document and partial replicas obtained.

Let's note $t_{\mathcal{V}_i} \le t_{\mathcal{V}_i}^{maj}$ the fact that the document $t_{\mathcal{V}_i}^{maj}$ is an update of the document $t_{\mathcal{V}_i}$, *i.e.* $t_{\mathcal{V}_i}^{maj}$ is obtained from $t_{\mathcal{V}_i}$ by replacing some of its buds by trees. In an asynchronous cooperative editing process, there are synchronization points[9] in which one tries to merge all contributions $t_{\mathcal{V}_i}^{maj}$ of the various co-authors to obtain a single comprehensive document $t_f$ [10]. A merging algorithm that does not incorporate conflict management and that relies on a solution to the *reverse projection* problem is given in [9].

### 2.2.2. Partial Replica and Reverse Projection (Expansion)

The *reverse projection* (also call the *expansion*) of an updated partial replica $t_{\mathcal{V}_i}^{maj}$ relatively to a given grammar $\mathbb{G} = (\mathcal{S}, \mathcal{P}, A)$, is the set $T_{iS}^{maj}$ of documents conform to $\mathbb{G}$ that admit $t_{\mathcal{V}_i}^{maj}$ as partial replica according to $\mathcal{V}_i$: $T_{iS}^{maj} = \left\{ t_{iS}^{maj} \therefore \mathbb{G} \mid \pi_{\mathcal{V}_i}\left(t_{iS}^{maj}\right) = t_{\mathcal{V}_i}^{maj} \right\}$.

A solution to this problem using tree automata is given in [9]; in that solution, productions of the grammar $\mathbb{G}$ are used, to bind to a view $\mathcal{V}_i \subseteq \mathcal{S}$ a tree automaton $\mathcal{A}^{(i)}$ such as the trees he recognizes from an initial state built from $t_{\mathcal{V}_i}^{maj}$ are exactly those having this partial replica as projection according to the view $\mathcal{V}_i$: $T_{iS}^{maj} = \mathcal{L}\left(\mathcal{A}^{(i)}, q_{t_{\mathcal{V}_i}}\right)$. Practically, a state $q$ of the automaton $\mathcal{A}^{(i)}$ is a pair $(Tag\ X, ts)$ where $X$ is a grammatical symbol, $ts$ is a forest (tree set), and $Tag$ is a label that is either *Open* or *Close* and indicates whether the concerned state $q$ can be used to generate a *closed* node or a *bud*. The states of $\mathcal{A}^{(i)}$ are typed: a state of the form $(Tag\ X, ts)$ is of type *X*. We also have a function named *typeState* which, when applied to a state returns its type[11]. A transition from one state $q$ is of one of the forms $(Close\ X, ts) \to (p, [q_1, \cdots, q_n])$ or $(Open\ X, [\,]) \to (X_\omega, [\,])$. A transition of the form $(Close\ X, ts) \to (p, [q_1, \cdots, q_n])$ is used to generate AST of type *X* (*i.e.* those whose root is labelled by a *X-production*) admitting "*ts*" for projection according to the view $\mathcal{V}_i$ if *X* does not belong to $\mathcal{V}_i$, and "$X[ts]$" otherwise. Similarly, a transition of the form $(Open\ X, [\,]) \to (X_\omega, [\,])$ is used to generate a single AST reduced to a bud of type *X*.

The interested reader may consult [9] for a more detailed description of the process of associating a tree automaton with a view and **Appendix A** for an illustration.

# 3. Reconciliation by Consensus

## 3.1. Issue and Principle of the Solution of Reconciliation by Consensus

There are generally two distinct phases when synchronizing replicas of a

[9]A synchronization point can be defined statically or triggered by a co-author as soon as certain properties are satisfied.

[10]It may happen that the edition must be continued after the merging (this is the case if there are still buds in the merged document): we must redistribute to each of the *n* co-authors a (partial) replica $t_{\mathcal{V}_i}$ of $t_f$ such that $t_{\mathcal{V}_i} = \pi_{\mathcal{V}_i}(t_f)$ for the continuation of the editing process.

[11] $typeState :: state \to symb$

$typeState(Open\ X, ts) = X$

$typeState(Close\ X, ts) = X$

document [14]: the *updates detection phase* which consists of recognizing the different replica nodes (locations) where updates have been made since the last synchronization, and the *propagation* phase which consists in combining the updates made on the various replicas to produce a new synchronized state (document) for each replica. In an asynchronous cooperative editing workflow of several *partial replicas* of a document, when you reach a synchronization point, you can end up with unmergeable replicas in their entirety as they contain not compatible updates[12] they must be reconciled. This can be done by questioning (cancelling) some local editing actions in order to resolve conflicts and result in a coherent global version said of consensus.

Studies on reconciling a document versions are based on heuristics [15] insofar as there is no general solution to this problem. In our case, since all editing actions are reversible[13] and it is easy to locate conflicts when trying to merge the partial replicas (see Section 3.2), we have a canonical method to eliminate conflicts: when merging, we replace any node (of the global document) whose replicas are in conflict by a bud. Thus, we prune at the nodes where a conflict appears, replacing the corresponding sub-tree with a bud of the appropriate type, indicating that this part of the document is not yet edited: the documents obtained are called consensus. These are the maximum prefixes without conflicts of the fusion of the documents resulting from the different expansions of the various updated partial replicas. For example, in **Figure 8**, the parts highlighted (blue backgrounds) in trees (f) and (g) are in conflict; they are replaced in the consensus tree (h) by a bud of type $C$ (node labelled $C_\omega$).

The problem of the consensual merging of $k$ updated partial replica whose global model is given by a grammar $\mathbb{G} = (S, \mathcal{P}, A)$ can therefore be stated as follows :

Problem of the consensual merging: Given $k$ views $(\mathcal{V}_i)_{1 \le i \le k}$ and $k$ partial replica $(t_{\mathcal{V}_i}^{maj})_{1 \le i \le k}$, merge consensually the family $(t_{\mathcal{V}_i}^{maj})_{1 \le i \le k}$ is to find the most large documents $t_S^{maj}$ conforming to $\mathbb{G}$ such that, for any document $t$ conforming to $\mathbb{G}$ and admitting $t_{\mathcal{V}_i}^{maj}$ as projection according to the view $\mathcal{V}_i$, $t_S^{maj}$ and $t$ are eventually updates each for other. *i.e.* (formula 2):

$$\left(t_S^{maj} \in \otimes_{i=1}^{k} t_i, t_i \in T_{iS}^{maj}\right) \Leftrightarrow \begin{cases} \text{i) } \forall i, 1 \le i \le k, \forall t \therefore \mathbb{G} \text{ such that } \pi_{\mathcal{V}_i}(t) = t_{\mathcal{V}_i}^{maj}, t_S^{maj} \cong t.^{14} \\ \text{ii) } \nexists t' \le t_S^{maj} \text{ such that } t' \in \otimes_{i=1}^{k} t_i, t_i \in T_{iS}^{maj} \end{cases} \quad (2)$$

The solution that we propose to this problem stems from an instrumentalization of that proposed for the expansion (Section 2.2.2). Indeed, we use an associative and commutative operator noted $\otimes$ to synchronize the tree automata $\mathcal{A}^{(i)}$ constructed to carry out the various expansions in order to generate the tree automaton of consensual merging. Noting $\mathcal{A}_{(sc)}$ this automaton, the documents of the consensus are the trees of the language generated by the automaton $\mathcal{A}_{(sc)}$ from an initial state built from the vector made of the initial states of the automata $\left(\mathcal{A}^{(i)}\right)$: $\mathcal{L}\left(\mathcal{A}_{(sc)},\left(q_{t^{maj}_{\mathcal{V}_i}}\right)\right) = consensus\left\{\mathcal{L}\left(\mathcal{A}^{(i)}, q_{t^{maj}_{\mathcal{V}_i}}\right)\right\}$.

[12]This is particularly the case if there is at least one node of the global document accessible by more than one co-author and edited by at least two of them using different productions.

[13]Reminder: the editing actions made on a partial replica may be cancelled as long as they do not have been incorporated into the global document.

[14]The binary relation $\cong$ when it exists between two trees $t_1$ and $t_2$ expresses the fact that they are possibly updates each for other. This relationship is more explicitly explained in definition 6.

$\mathcal{A}_{(sc)}$ is obtained by proceeding as follows: 1) For each view $\mathcal{V}_i$, build the automaton $\mathcal{A}^{(i)}$ who will carry out the expansion of the partial replica $t^{maj}_{\mathcal{V}_i}$ as previously indicated (Section 2.2.2): $\mathcal{L}\left(\mathcal{A}^{(i)}, q_{t^{maj}_{\mathcal{A}_i}}\right) = T^{maj}_{iS}$. 2) Using the operator $\otimes$, compute the automaton generating the consensus language $\mathcal{A}_{(sc)} = \otimes_{i=1}^{k} \mathcal{A}^{(i)}$.

## 3.2. Consensus Calculation

Before presenting the consensus calculation algorithm, let us specify using the concepts introduced in Section 2.1 the notion of (two) documents in conflict. Let $t_1, t_2 : \mathbb{N}^* \to \mathbf{A}$ be two trees (documents) with respectively $Dom(t_1)$ and $Dom(t_2)$ their domains. We say that $t_1$ and $t_2$ admit a consensus, and we note $t_1 \curlywedge t_2$, if theirs roots are of the same type[15] *i.e.* $(t_1 \curlywedge t_2) \Leftrightarrow \left(typeNode(t_1(\varepsilon)) = typeNode(t_2(\varepsilon))\right)$[16]. It is then say that $t_1$ and $t_2$ are in conflict, and it is noted $t_1 \curlyvee t_2$, when they admit a consensus but are not mergeable in their entirety. Intuitively, two documents $t_1$ and $t_2$ (not reduced to buds) are not fully mergeable (see Figure 6), if there exists an address $w \in Dom(t_1) \cap Dom(t_2)$ such that if we note $n_1$ (resp. $n_2$) the node located to address $w$ in $t_1$ (resp. in $t_2$), then, $n_1$ and $n_2$ which are not buds are of the same type but have different labels. *i.e.* (formula 3):

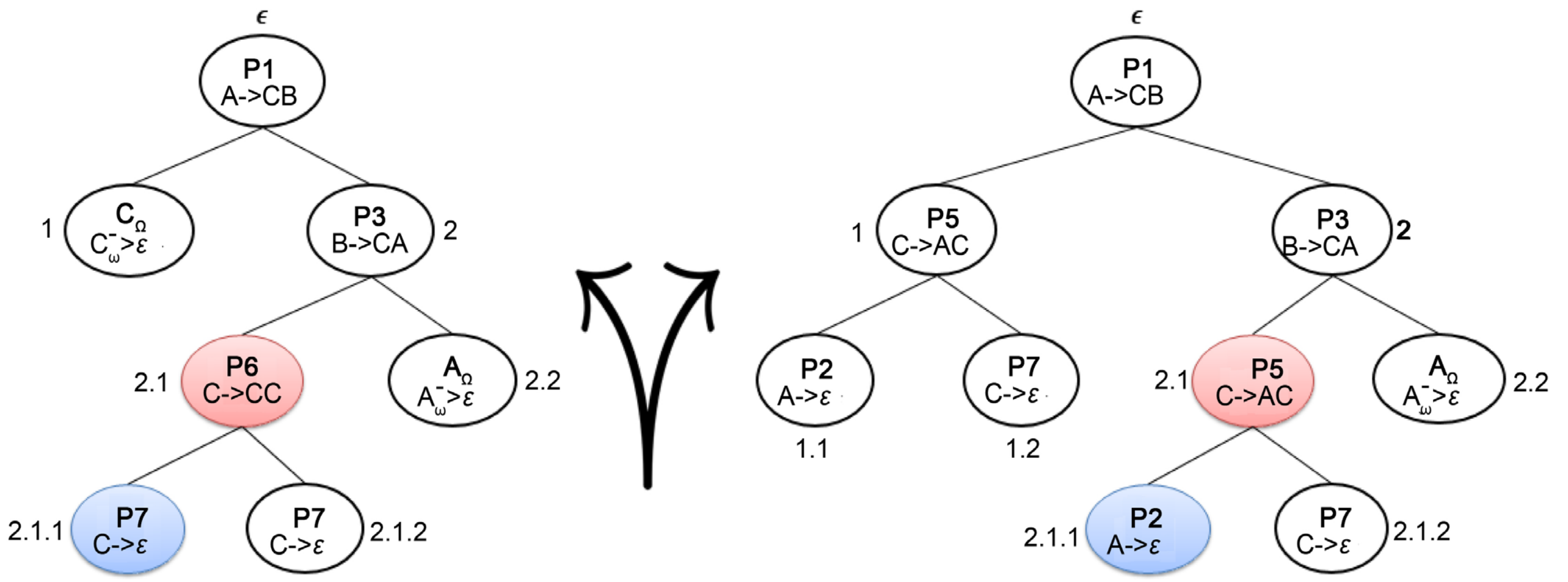

Figure 6. Example of documents in conflict.

[15]Trees we handle are AST and therefore, the nodes are labelled by productions names. Any node labelled by a *X-production* is said of *type X*. Furthermore, there is a function *type Node* such that *type Node* ($t(w)$) returns the type of the node located at the address $w$ in $t$.

[16]It may then be noted that two documents (AST) admit no consensus if their roots are of different types. However, for applications that interest us, namely structured editing, since the editions are done from the root (which is always of the type of the axiom) to the leaves using productions, the documents we manipulate always admit at least a consensus.

$$\left(t_1 \curlyvee t_2 \ with \ t_1(\varepsilon) \neq X_\omega, t_2(\varepsilon) \neq X_\omega\right) \Leftrightarrow \begin{cases} (t_1 \curlywedge t_2) \\ \text{and} \\ \left(\exists w \in Dom(t_1) \cup Dom(t_2), t_1(w) \neq t_2(w) \neq X_\omega \right. \\ \quad \left. \text{and } typeNode(t_1(w)) = typeNode(t_2(w))\right) \end{cases} \tag{3}$$

**Figure 6** shows two conflicting documents. In fact, at address 2.1 we have two nodes of the same type ("C") but edited with different C-productions: production $C \to C\,C$ in the first document, and production $C \to A\,C$ in the second one.

#### 3.2.1. Consensus among Multiple (Two) Documents

Let $t_1, t_2 : \mathbb{N}^* \to \mathbf{A}$ be two trees (documents) in conflict with respectively $Dom(t_1)$ and $Dom(t_2)$ their domains. The consensual tree $t_c : \mathbb{N}^* \to \mathbf{A}$ derived from $t_1$ and $t_2$ ($t_c = t_1 \otimes t_2$) has as domain the union of domains of the two trees in which we subtract elements belonging to domains of sub-trees derived from the conflicting nodes. In fact, we prune at the nodes in conflict and they appear in the consensus tree as a (unique) bud. So,

$$\forall w \in Dom(t_c), t_c(w) = \begin{cases} t_1(w) & \text{if } typeNode(t_1(w)) = typeNode(t_2(w)) \text{ and } t_1(w) = t_2(w) \\ t_1(w) & \text{if } typeNode(t_1(w)) = typeNode(t_2(w)) \text{ and } t_2(w) = X_\omega \\ t_2(w) & \text{if } typeNode(t_1(w)) = typeNode(t_2(w)) \text{ and } t_1(w) = X_\omega \\ t_1(w) & \text{if } w \notin Dom(t_2) \text{ and } \exists u, v \in \mathbb{N}^* \ tq \ w = u.v, \\ & \quad t_2(u) = X_\omega \text{ and } typeNode(t_1(u)) = typeNode(t_2(u)) \\ t_2(w) & \text{if } w \notin Dom(t_1) \text{ and } \exists u, v \in \mathbb{N}^* \ tq \ w = u.v, \\ & \quad t_1(u) = X_\omega \text{ and } typeNode(t_1(u)) = typeNode(t_2(u)) \\ X_\omega & \text{if } typeNode(t_1(w)) = typeNode(t_2(w)) \text{ and } t_1(w) \neq X_\omega \\ & \quad \text{and } t_2(w) \neq X_\omega \text{ and } t_1(w) \neq t_2(w) \end{cases} \tag{4}$$

**Figure 7** present the document resulting from the consensual merging of the documents in Figure 6. We have prume at the level of nodes 2.1 in both documents who are in conflict.

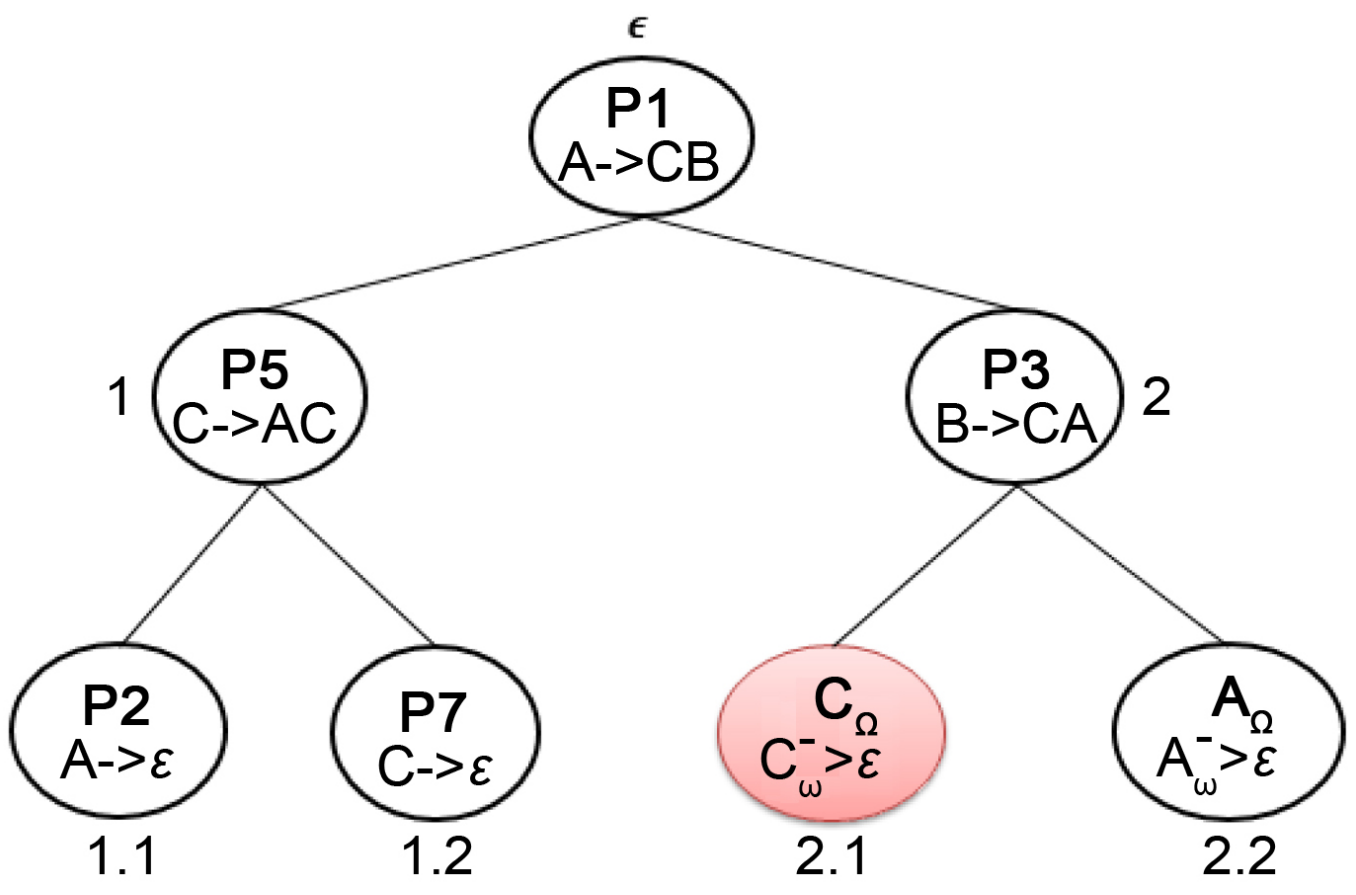

**Figure 7.** Document resulting from the consensual merging of the documents in **Figure 6**.

When $t_c = t_1 \otimes t_2$, there may be nodes of $t_1$ and those of $t_2$ which are updates of the nodes of $t_c$: it is said in this case that $t_1$ (resp. $t_2$) and $t_c$ are *updates each for other*.

**Definition 6.** *Let* $t_1, t_2$ *be two documents that are not in conflict. It will be said that they are updates each for other and it is noted* $t_1 \cong t_2$, *if there exists at least two addresses* $w, w'$ *of their respective domains such that* $t_1(w)$ *(resp.* $t_2(w')$*) is a bud and* $t_2(w)$ *(resp.* $t_1(w')$*) is not.*

### 3.2.2. Construction of Consensus Automaton

Consideration of documents with buds requires the readjustment of some models. For example, in the following, we will handle the *tree automata with exit states* instead of tree automata introduced in definition 4. Intuitively, a state $q$ of an automaton is called an *exit state* if there is a unique transition $q \to (X_\omega, [\,])$ associated to it for generating a tree reduced to a bud of type $X \in \Sigma$: $q$ is then of the form (*Open X*, []).

A tree automata with exit states $\mathcal{A}$ is a quintuplet $(\Sigma, Q, R, q_0, exit)$ where $\Sigma, Q, R, q_0$ designate the same objects as those introduced in the definition 4, and $exit$ is a predicate defined on the states ($exit : Q \to Bool$). Any state $q$ of $Q$ for which $exit\ q$ is $True$ is an exit state.

A type for automata with exit states can be defined in Haskell [16] [17] by:

```
data Automata prod state = Auto{
            exit:: state -> Bool,
            next :: state -> [(prod,[state])]
            }
```

In Section 3.2.1 above, we said that, *when two nodes are in conflict*, "*they appear in the consensus tree as a* (*unique*) *bud*". From the point of view of automata synchronization, the concept of "nodes in conflict" is the counterpart of the concept of "states in conflict" (as we specify below) and the above extract is reflected in the automata context by "*when two state are in conflict, they appear in the consensus automaton in the form of a* (*single*) *exit state*". Thus, if we consider two states of the same type $q_0^1$ and $q_0^2$ (which are not exit states) of two automata $auto_1$ and $auto_2$ with associated transitions families respectively $q_0^1 \to \left[\left(a_1^1, qs_1\right), \cdots, \left(a_{n_1}^1, qs_{n_1}\right)\right]$ and $q_0^2 \to \left[\left(a_1^2, qs_1'\right), \cdots, \left(a_{n_2}^1, qs_{n_2}'\right)\right]$, we say that the states $q_0^1$ and $q_0^2$ are in conflict (and we note $q_0^1 \curlyvee q_0^2$) if there is no transition starting from each of them and with the same label, *i.e.*

$$\left(q_0^1 \curlyvee q_0^2\right) \Leftrightarrow \left(\nexists a^3, \left(a^3, qs\right) \in \left\{next\ q_o^1\right\}, \left(a^3, qs'\right) \in \left\{next\ q_o^2\right\}, |qs| = |qs'|\right)$$

This can be coded in Haskell by the following function:

```
areInConflict state1@(tagsymb1, ts1) state2@(tagsymb2, ts2) =
               null [a1 | (a1,states1) <- next auto1 state1,
                          (a2,states2) <- next auto2 state2,
                           a1==a2,
                          (length states1)==(length states2)]
```

If $X$ is the type of two states $q$ and $q'$ in conflict, they admit a single consensual state $q_\omega = (Open\ X, [\,])$ such as $next\ q_\omega = [(X_\omega, [\,])]$. It is therefore obvious that two given automata admit a consensual automaton when their initials states are of the same type. The following function performs this test.

```
haveConsensus q0 q0′ = (typeState q0) == (typeState q0′)
```

The operator $\otimes$ used to calculate the synchronized consensual automaton $\mathcal{A}_{(sc)} = \otimes_i^k \mathcal{A}^{(i)}$ is a relaxation of the operator used for calculating the automata product presented in the definition 5. $\mathcal{A}_{(sc)} = (\Sigma, Q, R, q_0, exit)$ is an automaton with exit states and is constructed as follows:

- Its states are vectors of states : $Q = Q^{(1)} \times \cdots \times Q^{(k)}$;
- its initial state is formed by the vector of initial states of different automata: $q_0 = \left(q_0^{(1)}, \cdots, q_0^{(k)}\right)$;
- For the *exit* function , it is considered that when a given automaton $\mathcal{A}^{(j)}$ reached an *exit state*[17], it no longer contributes to behavior, but is not opposed to the synchronization of the others automata: it is said "*asleep*" (see listing "Consensus Listing" below, lines 16, 18 and 23). So, a state $q = (q^1, \cdots, q^k)$ is an exit state if: (a) all composite states $q^i$ are asleep (see listing "Consensus Listing" below, line 5) or (b) there exist any two states $q^i$ and $q^j, i \neq j$, components of $q$ that are in conflict (see listing "Consensus Listing" below, line 11)

$$\left(exit\left(q^{(1)}, \cdots, q^{(k)}\right)\right)$$
$$\Leftrightarrow \left(\left(exit\ q^{(i)}, \forall i \in \{1, \cdots, k\}\right) \text{ or } \left(\exists i, j, i \neq j, q^{(i)} \curlyvee q^{(j)}\right)\right)$$

- Its transitions are given by:

a) If $exit\ q$ then $q \to (X_\omega, [\,])$ is the unique transition of $q$; $X$ is the type of $q$.

b) Else $\left(q^{(1)}, \cdots, q^{(k)}\right) \xrightarrow{a} \left(\left(q_1^{(1)}, \cdots, q_1^{(k)}\right), \cdots, \left(q_n^{(1)}, \cdots, q_n^{(k)}\right)\right) \Leftrightarrow \forall i, 1 \leq i \leq k$

b1) $exit\ q^{(i)}$ and $\left(q_j^{(i)} = (Open\ X, [\,]), \forall j, 1 \leq j \leq n\right)$ /* $q^{(i)}$ *is asleep**/, else

b2) $q^{(i)} \xrightarrow{a} \left(q_1^{(i)}, \cdots, q_n^{(i)}\right)$.

a) reflects the fact that if a state $q$ is an exit one, we associate a single transition for generating a tree reduced to a bud of the type of $q$ (see listing "Consensus Listing" below, line 11).

With (b1) we say that, if the component $q^{(i)}$ of $q$ is an exit state, then for all composite state $\left(q_j^{(1)}, \cdots, q_j^{(k)}\right)$, ($1 \leq j \leq n$) appearing in the right hand side in the transition (b), the $i^{th}$ component should be asleep. Since it must not prevent other non-asleep states to synchronize, it must be of the form $(Open\ X, [\,])$ where $X$ is the type of the other states $q_j^{(l)}$ (being yet synchronized) belonging to $\left(q_j^{(1)},\cdots,q_j^{(k)}\right)$ (see function *forward Sleep State* defined in listing "Consensus Listing" below line 23 and used in lines 16 and 18). Finally, with (b2) we stipulate that if $q^{(i)} \xrightarrow{a} \left(q_1^{(i)},\cdots,q_n^{(i)}\right)$ is a transition of the automaton $\mathcal{A}^i$, then for all composite state $\left(q_j^{(1)},\cdots,q_j^{(k)}\right)$, ($1 \le j \le n$) appearing in the right part in the transition (b) above, the $i^{th}$ component is $q_j^{(i)}$ (see listing "Consensus Listing" below, lines 12 to 15).

[17]The corresponding node in the reverse projection of the document is a bud and reflects the fact that the corresponding author did not publish it. In the case that this node is shared with another co-author who published it in its (partial) replica, it is the edition made by the latter that will be retained when merging.

Consensus Listing

```
1  autoconsens::(Eq p, Eq x) =>(x -> p) -> Automata p (Tag x, [st1])
2       -> Automata p (Tag x, [st2]) -> Automata p ((Tag x, [st1]), (Tag x, [st2]))
3  autoconsens symb2prod auto1  auto2 = Auto exit_ next_ where
4    exit_ (state1,state2) = case haveConsensus state1 state2 of
5      True  -> (exit auto1 state1) && (exit auto2 state2)
6      False -> True
7    next_ (state1,state2) = case haveConsensus state1 state2 of
8      False -> []
9      True  -> case (exit auto1 state1, exit auto2 state2) of
10        (False,False) -> case (areInConflict state1 state2) of
11          True  -> [(symb2prod(typeState state1),[])]
12          False -> [(a1,zip states1 states2) |
13                   (a1,states1) <- next auto1 state1,
14                   (a2,states2) <- next auto2 state2,
15                   a1==a2, (length states1)==(length states2)]
16        (False,True)  -> [(a, zip states1 (forwardSleepState  states1)) |
17                         (a,states1) <- next auto1 state1]
18        (True,False)  -> [(a, zip (forwardSleepState  states2) states2) |
19                         (a,states2) <- next auto2 state2]
20        (True,True)   -> [(a1,[]) | (a1,[]) <- next auto1 state1,
21                         (a2,[]) <- next auto2 state2, a1==a2]
22   where
23     forwardSleepState states = map (\state -> (Open (typeState state), [])) states
```

**Proposition 7.** *The tree automaton* $\mathcal{A} = \otimes_{i=1}^{k}\mathcal{A}^{(i)}$ *recognizes/generates from the initial state* $q_0 = \left(q_{01},\cdots,q_{0k}\right)$ *all the trees from the consensual merging of trees recognized/generated by each automaton* $\mathcal{A}^{(i)}$ *from the initial state* $q_{0i}$. *Moreover, these trees are the biggest prefixes without conflicts of merged trees.*

$$\left(\otimes_{i=1}^{k}\mathcal{A}^{(i)} \vDash t \rhd q_0\right) \Leftrightarrow \begin{cases} \text{i) } \forall i \quad \exists t_i \quad \mathcal{A}^{(i)} \vDash t_i \rhd q_{0i} \text{ and } t_i \cong t \\ \text{ii) } \forall t' \text{ prefix of } t,\ \neg\left(\otimes_{i=1}^{k}\mathcal{A}^{(i)} \vDash t' \rhd q_0\right) \end{cases} \tag{5}$$

*Proof.* A tree *t* is recognized by the synchronized automaton $\otimes_{i=1}^{k}\mathcal{A}^{(i)}$ if, and only if, one can label each of its nodes by a state of the automaton in accordance with what is specified by the transitions of the automaton. Moreover, all the leaf nodes of *t* must be labelled by using *final transitions*; in our case, they are of the form $q \to \left(p,[\,]\right)$. This means that if a node whose initial label is *a* is labelled by the state *q* and if it admits *n* successors respectively labelled by $q_1,\cdots,q_n$, then $q \xrightarrow{a} \left(q_1,\cdots,q_n\right)$ must be a transition of the automaton. As the automaton is deterministic[18] this labelling is unique elsewhere (including the initial state attached to the root of the tree). By focusing our attention both on the state *q* labelling a node and its $i^{th}$ component $q_i$, on each of the branches of *t*, 1) we cut as soon as we reach an exit state in relation to the automaton $\mathcal{A}^{(i)}$ (*i.e.* $q_i$ is an exit state), or, 2) if $q$ is an exit state (in this case we are handling a leaf) and $q_i$ is not, relative to $\mathcal{A}^{(i)}$ (in this case, $q_i$ was in conflict with at least one other component $q_j$ of $q$); we replace that node with any sub-tree $t'_i$ that can be generated by $\mathcal{A}^{(i)}$ from the state $q_i$. So,

$$\left(\otimes_{i=1}^{k}\mathcal{A}^{(i)} \vDash t \rhd q_0\right) \Rightarrow \left(\forall i\ \exists t_i\quad \mathcal{A}^{(i)} \vDash t_i \rhd q_{0i} \text{ and } t_i \cong t\right) \tag{6}$$

since a state of $\mathcal{A}$ is an exit one if and only if each of its components is (in the $\mathcal{A}^i$) or if at least two of its components are in conflict.

[18]Automata $\mathcal{A}^{(i)}$ being deterministic (see proposition 3.3.3 of [18]), $\otimes_{i=1}^{k}\mathcal{A}^{(i)}$ is deterministic as synchronous product of deterministic automata.

Conversely, suppose $\mathcal{A}^{(i)} \vDash t_i \rhd q_{0i}$, by definition of the synchronized consensual automaton, we have $\otimes_{i=1}^{k}\mathcal{A}^{(i)} \vDash \otimes_{i=1}^{k} t_i \rhd (q_{01},\cdots,q_{0k})$. So overall

$$L\left(\otimes_{i=1}^{k}\mathcal{A}^{(i)}, q_0\right) = \left\{\otimes_{i=1}^{k} t_i \mid \mathcal{A}^{(i)} \vDash t_i \rhd q_{0i}\right\} \tag{7}$$

Suppose that $t$ is recognized by $\otimes_{i=1}^{k}\mathcal{A}^{(i)}$; thus there is a labelling of its nodes with the states of $\otimes_{i=1}^{k}\mathcal{A}^{(i)}$ and as the transitions used for the labelling of its leaves are final. Let $t_p$ be a prefix of $t$. Let us show that $t_p$ is not recognized by $\otimes_{i=1}^{k}\mathcal{A}^{(i)}$ using the fact that any labelling of $t_p$ has at least one leaf node labelled by a state that is not associated to a final transition. The labels associated to the nodes of $t_p$ are the same as those associated to the nodes of same addresses in $t$ because $t_p$ is a prefix of $t$ and $\otimes_{i=1}^{k}\mathcal{A}^{(i)}$ is deterministic. $t_p$ is obtained from $t$ by pruning some sub-trees of $t$; so naturally he has a (non-zero) number of leaf nodes that can be developed to obtain $t$. Let us choose a such node and call it $n_f$. Suppose that it is labelled $p$ and was associated with a state $q_f = (q_1,\cdots,q_k)$ when labelling $t$. The *p_transition* that permit to recognize $n_f$ is not a final transition. Indeed, $n_f$ has in $t$ $|p|$ sons whose labels can be suppose to be the states $q_{f_1},\cdots,qf_{|p|}$. This means according to the labelling process and considering the fact that $\otimes_{i=1}^{k}\mathcal{A}^{(i)}$ is deterministic, that the single transition used for labelling $n_f$ and of its $|p|$ sons is $q_f \xrightarrow{p} \left(q_{f_1},\cdots,q_{f_{|p|}}\right)$ which is not a final transition. Therefore, $t_p$ is not recognized by $\otimes_{i=1}^{k}\mathcal{A}^{(i)}$. □

### 3.3. Illustration

Figure 8 is an illustration of an asynchronous cooperative editing process generating partial replicas (Figure 8(c) and Figure 8(e)) in conflict[19] from the grammar having as productions:

$$\begin{array}{lll} P_1: A \to C\ B & P_3: B \to C\ A & P_5: C \to A\ C \\ P_2: A \to \varepsilon & P_4: B \to B\ B & P_6: C \to C\ C \\ & & P_7: C \to \varepsilon \end{array} \tag{8}$$

Initially in the process, two partial replicas (Figure 8(b) and Figure 8(d)) are obtained by projections of the global document (Figure 8(a)). After their update (Figure 8(c) and Figure 8(e)) a synchronization point is reached and, by applying the approach described in Section 3.1 *i.e*, association of tree automata $\mathcal{A}^{(1)}$ and $\mathcal{A}^{(2)}$ respectively to the partial replicas $tv1$ and $tv2$, their consensual synchronization in the automaton $\mathcal{A}_{(sc)} = \mathcal{A}^{(1)} \otimes \mathcal{A}^{(2)}$, and finally, generation of consensus trees (Figure 8(h)). Remember that this example is fully unfold in Appendix A: therein, we present the different manipulated automata and a set of the simplest consensus documents (Figure 9).

[19]By realising expansions of each of the replicas, we respectively obtain among others, the documents presented by Figure 8(f) and Figure 8(g) on which one can easily observe a conflict highlighted by areas having a blue background.

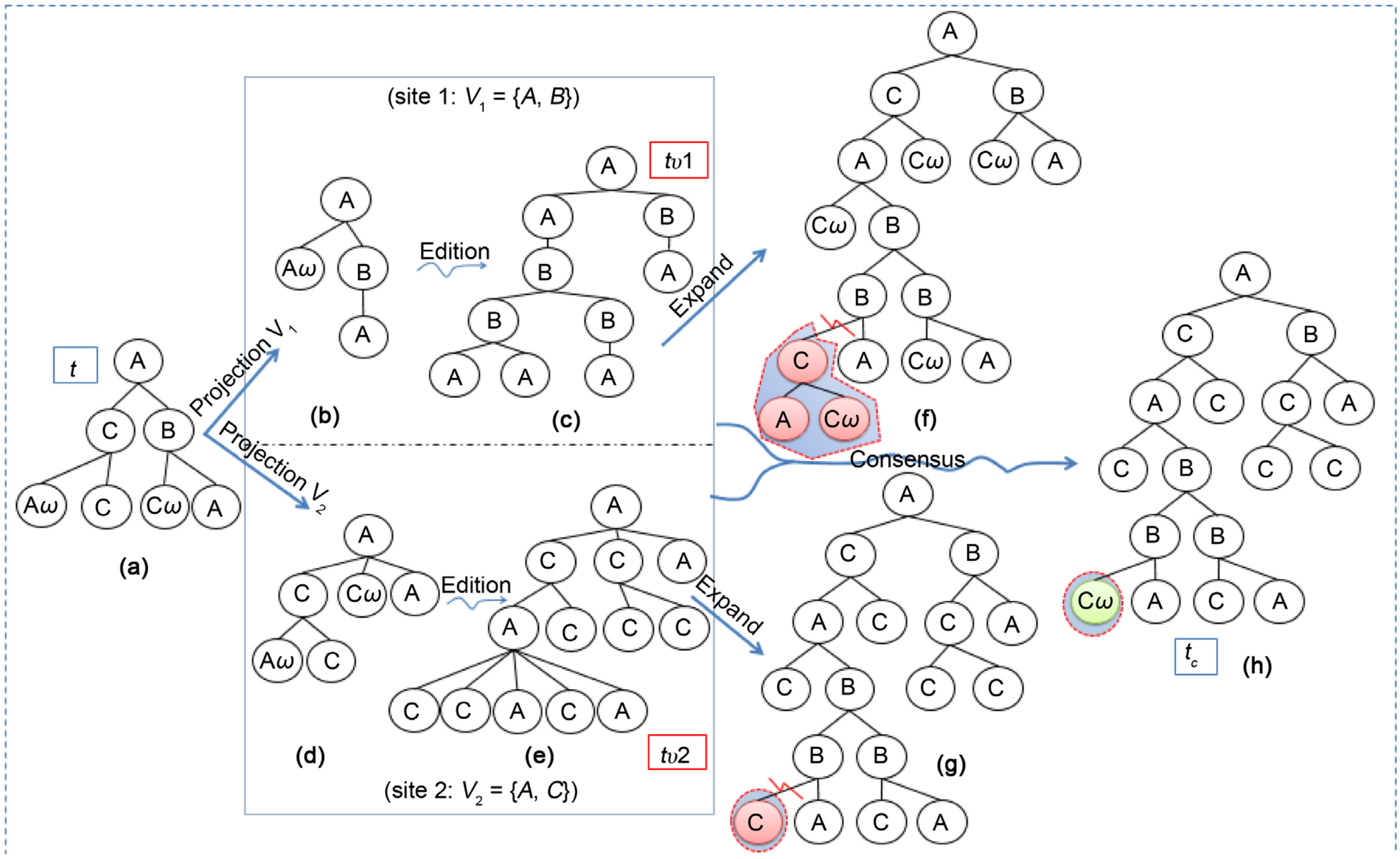

**Figure 8.** An edition with conflicts and corresponding consensus.

## 4. Conclusions

We presented in this paper a reconciliation approach said by *consensus*, of partial replicas of a document submitted to an asynchronous cooperative editing process: so we opted for a partial optimistic replication approach [12]. The approach proposed is based on a relaxation of the synchronous product of automata to construct an automaton capable of generating consensus documents.

The approach proposed in this paper is supported by mathematical proofs of the proposals. The presented algorithms have been implemented in Haskell [16] and experienced in many examples (including the one introduced in Section 3.3 and fully unfold in Appendix A) with convincing results. These algorithms can be also experienced in a truly distributed environment via the graphical editor prototype that we have built for this need; some screenshots are provided in Appendix B.

The deployment and use of this prototype will probably be better off if one incorporates a publishing environment generator which, from a specification of an asynchronous cooperative editing process describing in a DSL (Domain Specific Language) [19], the model of licit documents (grammar), various co-authors, their publishing sites and views, etc., will generate for each co-author her dedicated publishing environment including, for example, among others: a dedicated editor with conventional facilities of currents editors (syntax highlighting, code completion, etc.), tools for asking synchronizations, tools for backup and restoration of partial replicas being edited etc.

## Appendix A. An Illustration of the Merging Algorithm

We illustrate the consensual merging algorithm with the grammar of Section 3.3 (formula 8). We associate the automata $\mathcal{A}^{(1)}$ and $\mathcal{A}^{(2)}$ respectively to the updated partial replicas $tv1$ and $tv2$ (**Figure 8(c)** and **Figure 8(e)**), then we build the automaton of consensus $\mathcal{A}_{(sc)} = \mathcal{A}^{(1)} \otimes^{\Omega} \mathcal{A}^{(2)}$ by applying the approach described in Section 3.2.2 and finally, present the simplest documents of consensus (**Figure 9**).

### Linearization of a Structured Document

To simplify the presentation, we represent in the following, trees by their linearization in the form of a Dyck word. To do this, we associate a (various) pair of brackets to each grammatical symbol and the linearization of a tree is obtained by performing a Depth First Search (DFS) of the resulting tree.

### The Transition Schemas for the View {*A*, *B*}

A list of trees (forest) is represented by the concatenation of their linearizations.We use the opening parenthesis "('and the closing one')" to represent Dyck symbols associated with the visible symbol $A$ and the opening bracket "[" and closing "]" to represent those associated with the visible symbol $B$. Each transition of the automata associated to partial replicas according to the view $\{A,B\}$ is conform to one of the following transition schema[20]

$$\begin{array}{ll}
\langle A,w_1\rangle \to \left(P_1,\left[\langle C,u\rangle,\langle B,v\rangle\right]\right) & \text{if } w_1 = u[v] \\
\langle A,w_2\rangle \to \left(P_2,[\,]\right) & \text{if } w_2 = \varepsilon \\
\langle B,w_3\rangle \to \left(P_3,\left[\langle C,u\rangle,\langle A,v\rangle\right]\right) & \text{if } w_3 = u(v) \\
\langle B,w_4\rangle \to \left(P_4,\left[\langle B,u\rangle,\langle B,v\rangle\right]\right) & \text{if } w_4 = [u][v] \\
\langle C,w_5\rangle \to \left(P_5,\left[\langle A,u\rangle,\langle C,v\rangle\right]\right) & \text{if } w_5 = (u)v \\
\langle C,w_6\rangle \to \left(P_6,\left[\langle C,u\rangle,\langle C,v\rangle\right]\right) & \text{if } w_6 = uv \neq \varepsilon \\
\langle C,w_7\rangle \to \left(C_\omega,[\,]\right) & \text{if } w_7 = \varepsilon
\end{array} \quad (9)$$

These schemas are obtained from the grammar productions [9] and the pairs $\langle X,w_i\rangle$ are states where $X$ is a grammatical symbol and $w_i$ a forest encoded in the Dyck language. The first schema for example, states that the Abstract Syntax Tree (AST) generated from the state $\langle A,w_1\rangle$ are those obtained using the pro- duction $P_1$ to create a tree of the form $P_1[t_1,t_2]$; $t_1$ and $t_2$ being generated respectively from the states $\langle C,u\rangle$ and $\langle B,v\rangle$ such that $w_1 = u[v]$. The state $\langle C,w_7\rangle$ with $w_7=\varepsilon$ being an exit state [9], the rule $\langle C,w_7\rangle \to (C_\omega,[\,])$ linked to the production $P_7$ states that the AST generated from the state $\langle C,w_7\rangle$ is reduced to a bud of type $C$ ($C$ is the symbol located at the left hand side of $P_7$).

[20]We do not represent the whole set of transition schemas in this example; only the useful subset for reconciliation of closed documents is shown here because the documents to reconcile in this example are all closed (has no buds). To consider buds, one should, for each visible sort $X$, associate a new pair of Dyck symbols to the bud of type $X$ then, derive the new schemas.

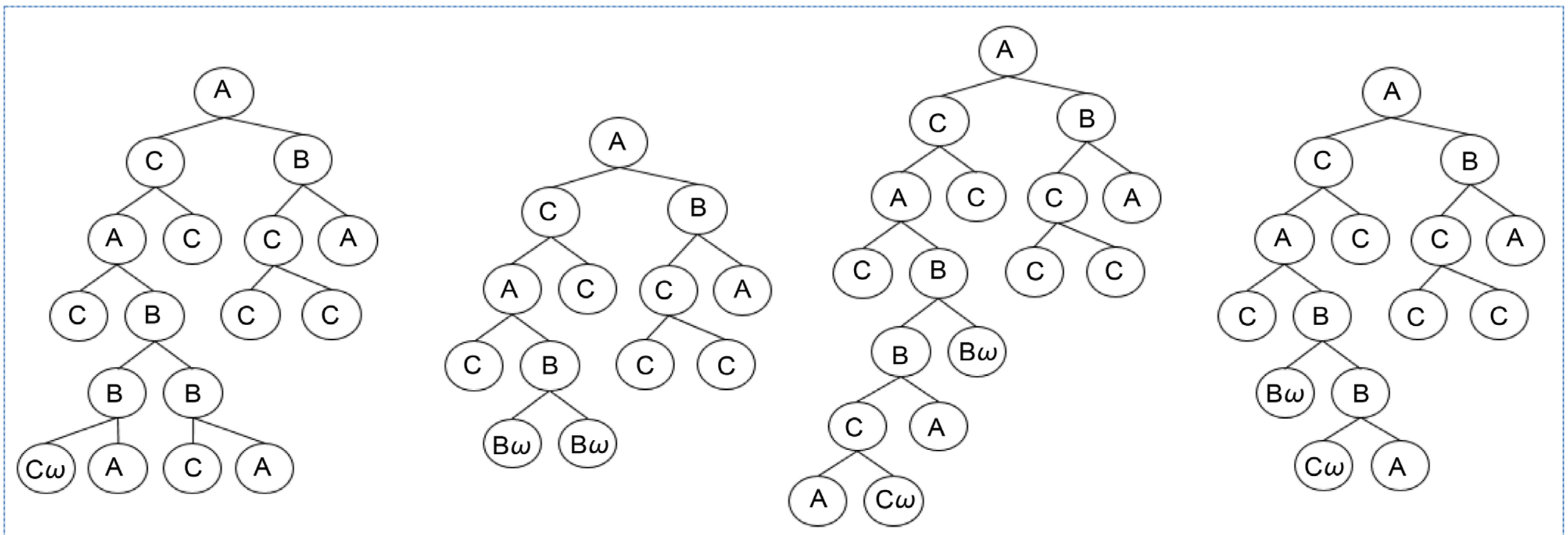

**Figure 9.** Consensual trees generated from the automaton $\mathcal{A}_{(sc)}$.

## Construction of the Automaton $\mathcal{A}^{(1)}$ Associated to *tv*1

Having associated Dyck symbols "('and')" (resp. "['and']") to the grammatical symbol $A$ (resp. $B$), the linearization of the partial replica $tv1$ (**Figure 8(c)**) gives $(([[()()][()]])[()])$. As $A$ is the axiom of the grammar, the initial state of the automaton $\mathcal{A}^{(1)}$ is $q_0^1 = \langle A, ([[()()][()]])[()] \rangle$. When considering only the states accessible from $q_0^1$ and by applying the previous schema of transition, we obtain the following automaton (**Table 1**) for the replica $tv1$ (**Figure 8(c)**).

The state $q_4^1 = \langle C, \varepsilon \rangle$ in **Table 1** is the only exit state of $\mathcal{A}^{(1)}$. It is easy to verify that the document of **Figure 8(f)** resulting from the reverse projection of $tv1$ belongs to the language accepted by the automaton $\mathcal{A}^{(1)}$.

## Construction of the Automaton $\mathcal{A}^{(2)}$ Associated to *tv*2

As before, by associating to the grammatical symbol $C$ (resp. $A$) the Dyck symbols "['and']" (resp. "('and')"), we obtain the transition schemas (formula 10) for the automata associated to the partial replicas according to the view $\{A, C\}$.

$$
\begin{aligned}
&\langle A, w_1 \rangle \to \left(P_1, \left[\langle C, u \rangle, \langle B, v \rangle\right]\right) && \text{if } w_1 = [u]v \\
&\langle A, w_2 \rangle \to \left(P_2, [\,]\right) && \text{if } w_2 = \varepsilon \\
&\langle B, w_3 \rangle \to \left(P_3, \left[\langle C, u \rangle, \langle A, v \rangle\right]\right) && \text{if } w_3 = [u](v) \\
&\langle B, w_4 \rangle \to \left(P_4, \left[\langle B, u \rangle, \langle B, v \rangle\right]\right) && \text{if } w_4 = uv \neq \varepsilon \\
&\langle B, w_5 \rangle \to \left(B_\omega, [\,]\right) && \text{if } w_5 = \varepsilon \\
&\langle C, w_6 \rangle \to \left(P_5, \left[\langle A, u \rangle, \langle C, v \rangle\right]\right) && \text{if } w_6 = (u)[v] \\
&\langle C, w_7 \rangle \to \left(P_6, \left[\langle C, u \rangle, \langle C, v \rangle\right]\right) && \text{if } w_7 = [u][v] \\
&\langle C, w_8 \rangle \to \left(P_7, [\,]\right) && \text{if } w_8 = \varepsilon
\end{aligned}
\tag{10}
$$

The linearization of the partial replica $tv2$ (**Figure 8(e)**) is $([([][]()[]())[]][[][]]())$. The automaton $\mathcal{A}^{(2)}$ associates to this replica has as initial state $q_0^2 = \langle A, [([][]()[]())[]][[][]]() \rangle$ and its transitions are given in Table 2.

**Table 1.** Automaton accepting updates of the (partial) replica *tv*1.

| | |
|---|---|
| $q_0^1 \rightarrow (P_1, [q_1^1, q_2^1])$ | with $q_1^1 = \langle C, ([[()()][()]]) \rangle$ and $q_2^1 = \langle B, () \rangle$ |
| $q_1^1 \rightarrow (P_5, [q_3^1, q_4^1])$ | with $q_3^1 = \langle A, [[()()][()]] \rangle$ and $q_4^1 = \langle C, \varepsilon \rangle$ |
| $q_1^1 \rightarrow (P_6, [q_4^1, q_1^1]) \mid (P_6, [q_1^1, q_4^1]])$ | |
| $q_2^1 \rightarrow (P_3, [q_4^1, q_5^1])$ | with $q_5^1 = \langle A, \varepsilon \rangle$ |
| $q_3^1 \rightarrow (P_1, [q_4^1, q_6^1])$ | with $q_6^1 = \langle B, [()()][()] \rangle$ |
| $q_4^1 \rightarrow (C_\omega, [])$ | |
| $q_5^1 \rightarrow (P_2, [])$ | |
| $q_6^1 \rightarrow (P_4, [q_7^1, q_2^1])$ | with $q_7^1 = \langle B, ()() \rangle$ |
| $q_7^1 \rightarrow (P_3, [q_8^1, q_5^1])$ | with $q_8^1 = \langle C, () \rangle$ |
| $q_8^1 \rightarrow (P_5, [q_5^1, q_4^1])$ | |
| $q_8^1 \rightarrow (P_6, [q_8^1, q_4^1]) \mid (P_6, [q_4^1, q_8^1])$ | |

Let's note that, the state $q_8^2 = \langle B, \varepsilon \rangle$ is the only exit state of the automaton $\mathcal{A}^{(2)}$.

### Construction of the Consensus Automaton $\mathcal{A}_{(sc)}$

By application of synchronous product of several tree automata described in Section 3.2.2 to the automata $\mathcal{A}^{(1)}$ and $\mathcal{A}^{(2)}$, the consensual automaton $\mathcal{A}_{(sc)} = \mathcal{A}^{(1)} \otimes^{\Omega} \mathcal{A}^{(2)}$ has $q_0 = (q_0^1, q_0^2)$ as initial state. $\mathcal{A}^{(1)}$ has a transition from $q_0^1$ to $[q_1^1, q_2^1]$ labelled $P_1$. Similarly, $\mathcal{A}^{(2)}$ has a transition from $q_0^2$ to $[q_1^2, q_2^2]$ labelled $P_1$. So we have in $\mathcal{A}_{(sc)}$ a transition labelled $P_1$ for accessing states $[q_1 = (q_1^1, q_1^2), q_2 = (q_2^1, q_2^2)]$ from the initial state $q_0 = (q_0^1, q_0^2)$. Following this principle, we construct the following consensual automaton (Table 3).

The states $\{q_{10}, q_{11}, q_{12}, q_{15}, q_{16}, q_{17}, q_{21}, q_{22}\}$ are the exit states of the automaton $\mathcal{A}_{(sc)}$. They are states whose composite states are either in conflict (for example $q_{10} = (q_2^1, q_7^2)$ et $q_2^1 q_7^2$), or are all exit states (for example $q_{22} = (q_4^1, q_{s1})$).

The use of function that generates the simplest AST (with buds) of a tree language from its automaton [9] on $\mathcal{A}_{(sc)}$, produced *four* AST whose derivation trees (the consensus) are shown schematically in Figure 9.

## Appendix B. A Prototype of a Cooperative Editor Using Our Algorithms

We present below some screenshots of the cooperative editor prototype for graphic and cooperative editing of the abstract structure of structured do-cuments using our algorithms for consensus merging of edited partial replicas.

Table 2. Automaton accepting updates of the (partial) replica *tv2*.

| | |
|---|---|
| $q_0^2 \to \left(P_1,\left[q_1^2,q_2^2\right]\right)$ | with $q_1^2 = \langle C,([][]()[]())[]\rangle$ and $q_2^2 = \langle B,[[][]]()\rangle$ |
| $q_1^2 \to \left(P_5,\left[q_3^2,q_4^2\right]\right)$ | with $q_3^2 = \langle A,[][]()[]()\rangle$ and $q_4^2 = \langle C,\varepsilon\rangle$ |
| $q_2^2 \to \left(P_3,\left[q_5^2,q_6^2\right]\right)$ | with $q_5^2 = \langle C,[][]\rangle$ and $q_6^2 = \langle A,\varepsilon\rangle$ |
| $q_3^2 \to \left(P_1,\left[q_4^2,q_7^2\right]\right)$ | with $q_7^2 = \langle B,[]()[]()\rangle$ |
| $q_4^2 \to (P_7,[])$ | |
| $q_5^2 \to \left(P_6,\left[q_4^2,q_4^2\right]\right)$ | |
| $q_6^2 \to (P_2,[])$ | |
| $q_7^2 \to \left(P_4,\left[q_8^2,q_7^2\right]\right) \mid \left(P_4,\left[q_9^2,q_{10}^2\right]\right) \mid \left(P_4,\left[q_{11}^2,q_{11}^2\right]\right) \mid \left(P_4,\left[q_{12}^2,q_{13}^2\right]\right) \mid \left(P_4,\left[q_7^2,q_8^2\right]\right)$ | with $q_8^2 = \langle B,\varepsilon\rangle$, $q_9^2 = \langle B,[]\rangle$, $q_{10}^2 = \langle B,()[]()\rangle$, $q_{11}^2 = \langle B,[]()\rangle$, $q_{12}^2 = \langle B,[]()[]\rangle$ and $q_{13}^2 = \langle B,()\rangle$ |
| $q_8^2 \to (B_\omega,[])$ | |
| $q_9^2 \to \left(P_4,\left[q_8^2,q_9^2\right]\right) \mid \left(P_4,\left[q_9^2,q_8^2\right]\right)$ | |
| $q_{10}^2 \to \left(P_4,\left[q_8^2,q_{10}^2\right]\right) \mid \left(P_4,\left[q_{13}^2,q_{11}^2\right]\right) \mid \left(P_4,\left[q_{14}^2,q_{13}^2\right]\right) \mid \left(P_4,\left[q_{10}^2,q_8^2\right]\right)$ | with $q_{14}^2 = \langle B,()[]\rangle$ |
| $q_{11}^2 \to \left(P_3,\left[q_4^2,q_6^2\right]\right)$ | |
| $q_{12}^2 \to \left(P_4,\left[q_8^2,q_{12}^2\right]\right) \mid \left(P_4,\left[q_9^2,q_{14}^2\right]\right) \mid \left(P_4,\left[q_{11}^2,q_9^2\right]\right) \mid \left(P_4,\left[q_{12}^2,q_8^2\right]\right)$ | |
| $q_{13}^2 \to \left(P_4,\left[q_8^2,q_{13}^2\right]\right) \mid \left(P_4,\left[q_{13}^2,q_8^2\right]\right)$ | |
| $q_{14}^2 \to \left(P_4,\left[q_8^2,q_{14}^2\right]\right) \mid \left(P_4,\left[q_{13}^2,q_9^2\right]\right) \mid \left(P_4,\left[q_{14}^2,q_8^2\right]\right)$ | |

This prototype is used following a networked client-server model. Its user interface offers to the user facilities for creating workflows: grammars, actors and views, initial document, … (Figure 10), editing and validation of partial replicas, connecting to a local or remote workflow (Figure 11). Moreover, this interface also offers him functionality to experience the concepts of projection, expansion and consensual merging (Figure 12). This prototype is designed using Java and Haskell languages.

**Table 3.** The consensus automaton.

| | |
|---|---|
| | $q_0 = (q_0^1, q_0^2)$ |
| $q_0 \to (P_1, [q_1, q_2])$ | with $q_1 = (q_1^1, q_1^2)$ and $q_2 = (q_2^1, q_2^2)$ |
| $q_1 \to (P_5, [q_3, q_4])$ | with $q_3 = (q_3^1, q_3^2)$ and $q_4 = (q_4^1, q_4^2)$ |
| $q_2 \to (P_3, [q_5, q_6])$ | with $q_5 = (q_4^1, q_5^2)$ and $q_6 = (q_5^1, q_6^2)$ |
| $q_3 \to (P_1, [q_4, q_7])$ | with $q_7 = (q_6^1, q_7^2)$ |
| $q_4 \to (P_7, [])$ | |
| $q_5 \to (P_6, [q_8, q_8])$ | with $q_8 = (q_{s1}, q_4^2)$ and $q_{s1} = \langle Open\ C, [] \rangle$ |
| $q_6 \to (P_2, [])$ | |
| $q_7 \to (P_4, [q_9, q_{10}]) \mid (P_4, [q_{11}, q_{12}]) \mid (P_4, [q_{13}, q_{14}]) \mid (P_4, [q_{15}, q_{16}]) \mid (P_4, [q_{17}, q_{18}])$ | with $q_9 = (q_7^1, q_8^2)$, $q_{10} = (q_2^1, q_7^2)$, $q_{11} = (q_7^1, q_9^2)$, $q_{12} = (q_2^1, q_{10}^2)$, $q_{13} = (q_7^1, q_{11}^2)$, $q_{14} = (q_2^1, q_{11}^2)$, $q_{15} = (q_7^1, q_{12}^2)$, $q_{16} = (q_2^1, q_{13}^2)$, $q_{17} = (q_7^1, q_7^2)$ and $q_{18} = (q_2^1, q_8^2)$ |
| $q_8 \to (P_7, [])$ | |
| $q_9 \to (P_3, [q_{19}, q_{20}])$ | with $q_{19} = (q_8^1, q_{s1})$ and $q_{20} = (q_5^1, q_{s2})$, $q_{s2} = \langle Open\ A, [] \rangle$ |
| $q_{13} \to (P_3, [q_{21}, q_6])$ | with $q_{21} = (q_8^1, q_4^2)$ |
| $q_{14} \to (P_3, [q_4, q_6])$ | |
| $q_{18} \to (P_3, [q_{22}, q_{20}])$ | with $q_{22} = (q_4^1, q_{s1})$ |
| $q_{19} \to (P_5, [q_{20}, q_{22}]) \mid (P_6, [q_{19}, q_{22}]) \mid (P_6, [q_{22}, q_{19}])$ | |
| $q_{20} \to (P_2, [])$ | |
| $q_{10} \to (B_\omega, [])$, $q_{11} \to (B_\omega, [])$, | |
| $q_{12} \to (B_\omega, [])$, $q_{15} \to (B_\omega, [])$, | |
| $q_{16} \to (B_\omega, [])$, $q_{17} \to (B_\omega, [])$, | |
| $q_{21} \to (C_\omega, [])$, $q_{22} \to (C_\omega, [])$ | |

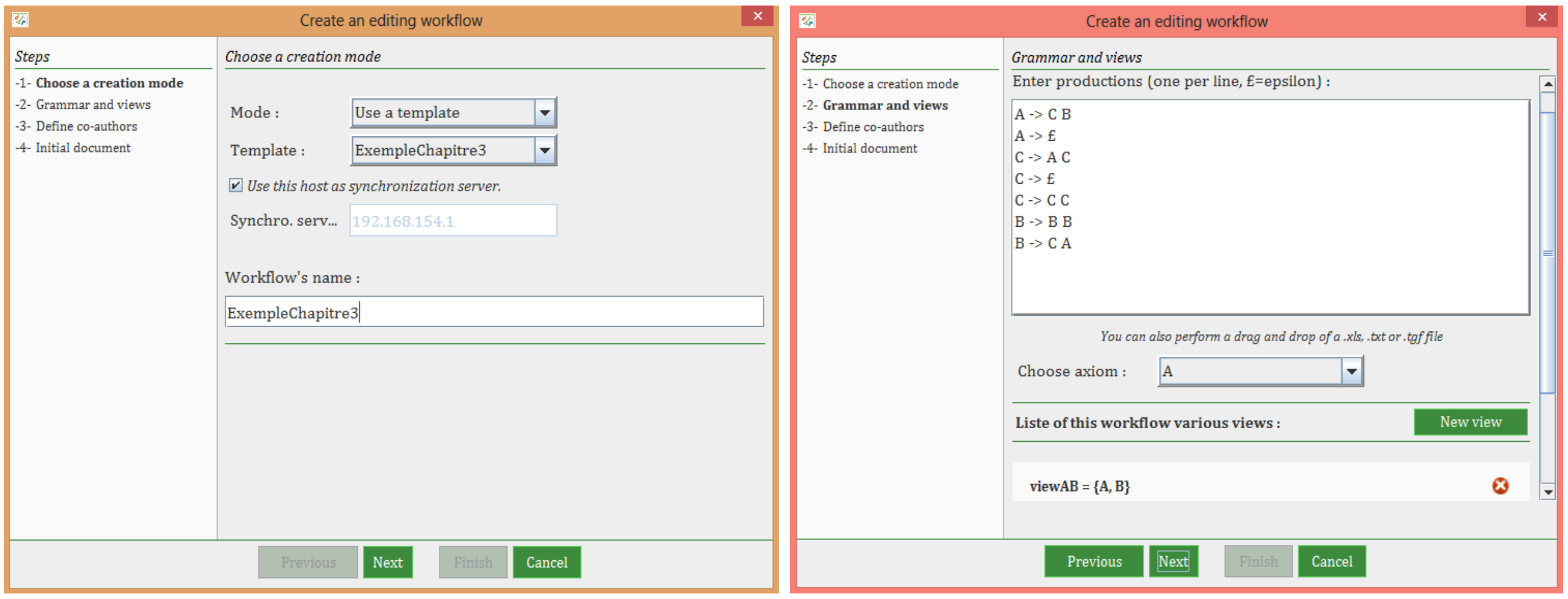

1- Workflow name, synchronization server and then, click on "Next"

2- Definitions of productions, axiom of the grammar and various views.

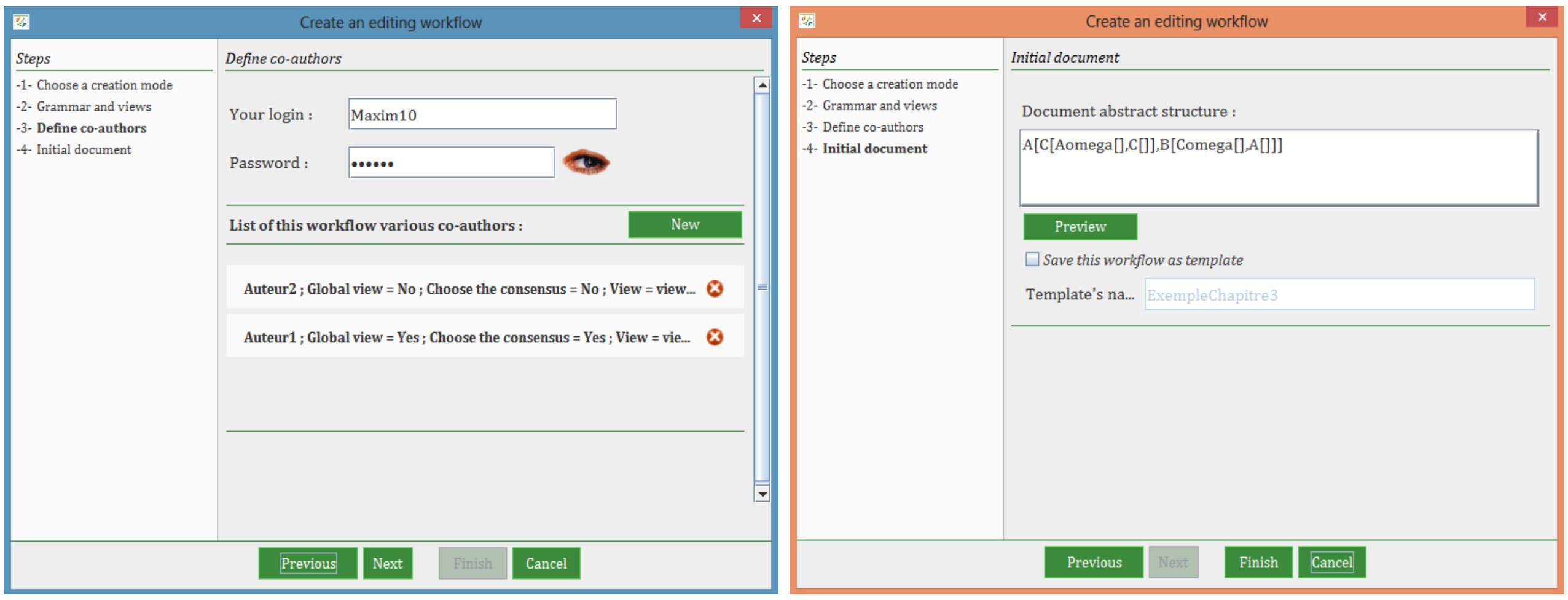

3- Informations on workflow owner and on different co-authors

4- Creation of the global initial document

**Figure 10.** Some prototype screenshots showing windows for the creation of a cooperative editing workflow.

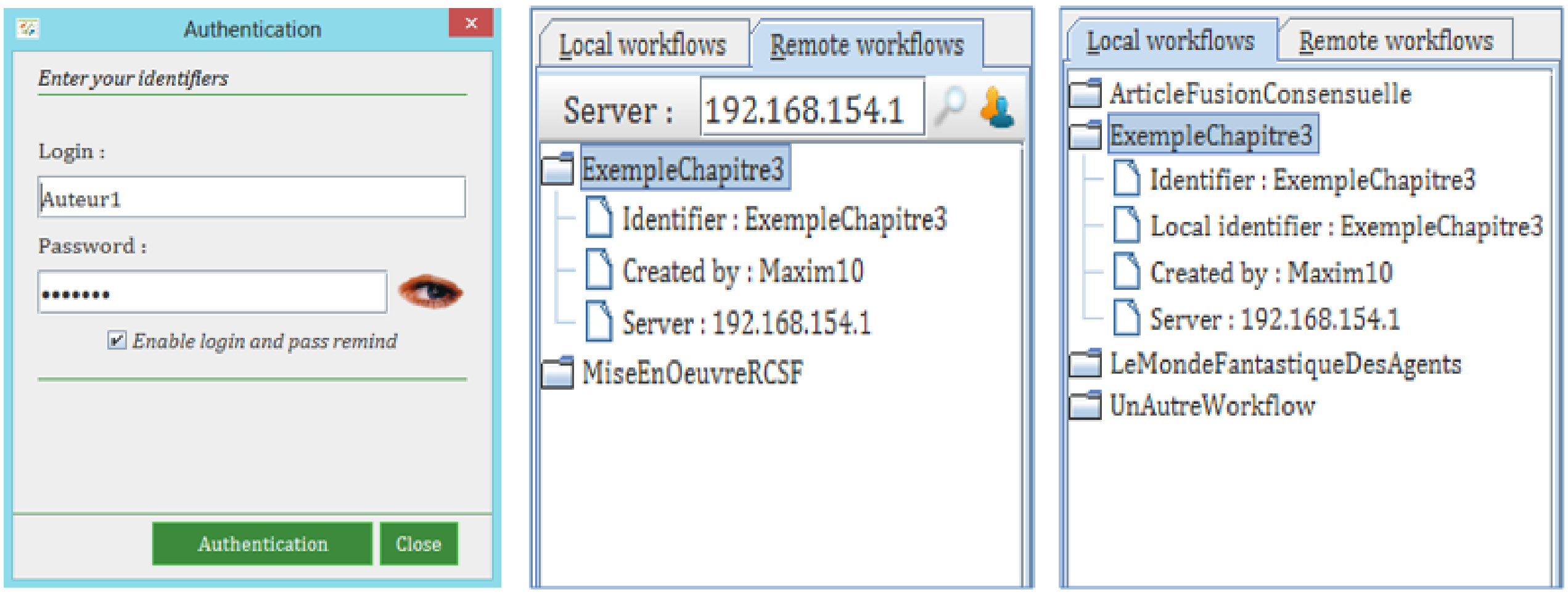

1- Authentication of the co-author Auteur1

2- Showing available remote workflows for Auteur1

3- Displaying local workflows of Auteur1

**Figure 11.** Some prototype screenshots showing the authentication window of a co-author (Auteur1) as well as those displaying the different workflows, local and remote in which he is implicated.

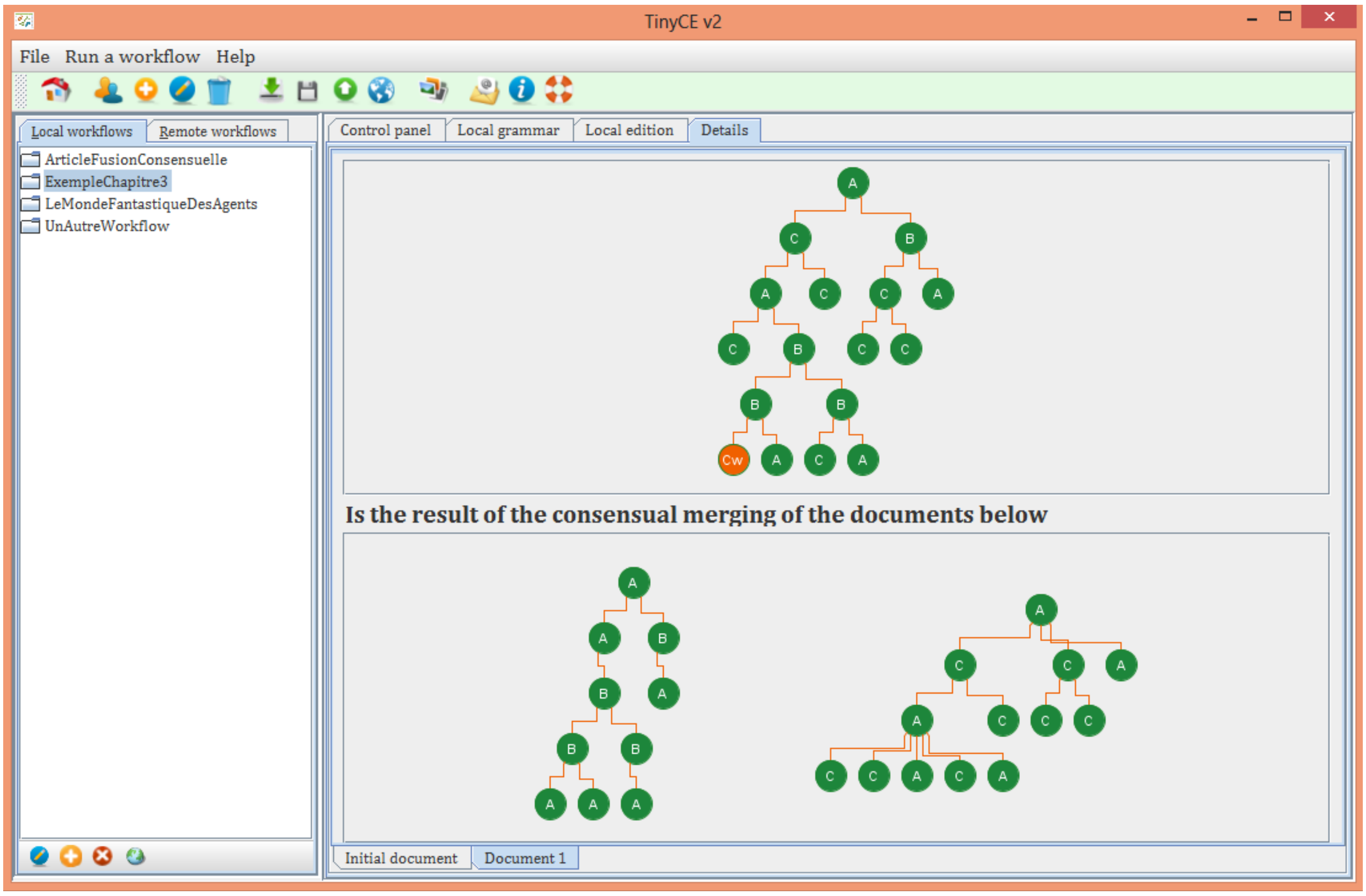

Figure 12. An illustration of the consensual merging in the prototype.